\documentclass[a4paper,11pt]{article}
\usepackage{graphicx}
\usepackage{dcolumn}
\newcommand{\beq}{\begin{equation}}
\newcommand{\eeq}{\end{equation}}
\newcommand{\Rns}{\textbf{R}$^*$}
\newcommand{\Rs}{\textbf{R}}
\newcommand{\Nns}{\textbf{N}$^*$}

\newcommand{\kk}{\mathbf{k}}

\def\tt{~}
\def\mdPi{monad$_{\rm \Pi}$ }
\title{A realistic model for completing Quantum Mechanics}
\author{M. Baldo}
\date{ }

\begin{document}
\maketitle
\vspace {-1 cm}
\begin{center}
\textit{Istituto Nazionale di Fisica Nucleare}
\par \textit{via Santa Sofia 64, 95123 Catania, Italy} 
\end{center}








\vskip 0.5 cm
\noindent
Abstract
\par
In the well known Copenhagen interpretation of Quantum mechanics, advocated by N. Bohr, the physical objects and the experimental results can be described only in a macroscopic language, leaving any possible microscopic description as 'unspeakable'. This point of view has been deepened by C. Rovelli in the 'relational interpretation' of Quantum mechanics. Most of the alternative interpretations, which try a detailed microscopic description of physical phenomena and of their evolution, have in common the explicit introduction
of the wave function as the basic element of the theory. These interpretations require the notion of 'quantum state' as
the fundamental concept of the theory, which is the typical 'unspeakable' physical element according to the Copenhagen interpretation. The two basic physical entities are intimately bound together by the integrity of the wave function. These interpretations are usually indicated as ''realistic". 
It is well known that the use of the wave function and its time evolution in the description of the physical processes leads unavoidably to some difficulties or so-called 'paradoxes'. The measurement problem is at the center of these difficulties, mainly because it requires the introduction of the reduction process of the wave function, which is not included explicitly within the mathematical formalism of Quantum Mechanics.\par
In this paper we build up and propose a model which goes beyond the standard formalism and which is able to solve the measurement problem and all the other difficulties which, in a way or in another, are related to it.


\par 

\section{Introduction}
\subsection{The main questions.\label{sec:questions}}
In this subsection we list and discuss the main problems and questions that will be considered. They refer to well known
facts and observations, but it is convenient to briefly review them to focus the purpose of the paper and its framework.
It will be shown that the model proposed in the paper is able to give an answer to each one of these questions. To the knowledge of the author no other model can solve all the questions listed below within the same unified scheme.
\vskip 0.3 cm
\par\noindent
{\it The reduction process.} 
\vskip 0.3 cm
\par 
In the Copenhagen interpretation of Quantum mechanics, advocated by N. Bohr \cite{Cope}, one cannot speak of the wave function reduction, since the concept itself of wave function is considered only as an additional devise to calculate the experimental results. This point of view has been extended in the 'relational interpretation' of Quantum Mechanics by C. Rovelli \cite{Rovelli1,Rovelli2,Rovelli3,Rovelli4}. Here we consider the alternative view, where the wave function and the corresponding  ''quantum state" are explicitly introduced as the basic element of the theory, while, in the terminology introduced by J. Bell \cite{Bell}, they are treated as  ''unspeakable" in the Copenhagen interpretation.
Let us first remind what the wave function reduction means.
In a typical measurement of a given physical quantity, i.e. 'observable', if the initial wave function of a microscopic physical system is a superposition of its eigenstates, it is assumed that the wave function after a single measurement by a proper detector reduces to a single eigenstate wave function. More in general, the detector plus the system will evolve to a single response among the different ones that correspond to the superposition.
This is the so-called reduction process, as presented in any textbook on Quantum Mechanics (QM) \cite{Berm,Bal,Peres}. To be more precise, this description corresponds to a 'strong measurement', as considered by von Neumann \cite{Neumann}. \par
The difficulties arise when one realizes that the reduction is necessarily a non linear process, since it breaks the initial superposition, which should be preserved if the process is governed by a linear evolution equation like the Schrodinger equation. This implies that to describe the reduction process one needs to go beyond standard QM.
One then introduces the Born's rule, which interprets the modulus square of each component in the initial wave function  as statistical probability to obtain the corresponding eigenvalue. Here an element of stochastic evolution is assumed to be present in the measurement process.\par 
The reduction and its stochastic character is demanded by observations, but it is a process not explicitly present in the QM mathematical formalism, which therefore must be considered as an additional postulate.
It has been proposed \cite{Joos,Schloss1,Schloss2,Schloss3,Olli1,Olli2,Blum1,Blum2,Blum3,Zur1,Zur2,Zur3} that the environment that surrounds the measurement apparatus could play a role. It can be argued that the interaction with the environment can destroy the coherence present in the initial state. In other words, the quantum density matrix pertinent to the initial state of the overall system could be reduced to a classical density matrix, where the off-diagonal elements are averaged to zero. However, a classical density matrix has still a statistical meaning, which indicates only the probability of each possible output in a set of measurements. The selection of a single value of the observable in a single measurement process remains still out of the formalism \cite{Schloss1,unsolv}. The fact that the different results are of course mutually exclusive does not imply the selection process.\par 
Indeterminacy is also a trivial consequence of the fact that different results can be obtained with the same initial state.\par 
Finally, the other essential property of the measurement process is irreversibility. Indeed the same output in a single measurement can be obtained with different initial wave functions, which implies the impossibility of reconstructing by time reversal the initial state from the result of a single measurement.\par 
In summary, the observed properties of measurement processes demand for an extension of the QM formalism once it uses explicitly the quantum state and the corresponding wave function as basic elements of the theory.
\vskip 0.3 cm
\par\noindent
{\it The classical limit.}
\vskip 0.3 cm
\par 
Since the structure of macroscopic classical objects has its basis on the microscopic realm, where QM is supposed to be valid, as the size of a physical system increases its description in terms of the QM formalism must merges into the classical one. This merging can be approximate, but very precise, or a definite separation between classical and quantum description does exist. The 'correspondence principle' \cite{corres} states that the classical limit must anyhow be valid. 
In a slightly restricted form, the principle assumes that for large quantum numbers the predictions of QM and classical physics must coincide to an arbitrary accuracy. This can be explicitly verified in numerous examples, and a general verification can be obtained in the path integral formulation of QM \cite{Feynman}, where the formal limit of the Planck' s constant going to zero can be more easily performed.\par 
 However it is well known that the evolution of a quantum state, initially prepared as a wave packet which is close to a classical state (in arbitrary phase space dimensions), can remain close to the corresponding classical trajectory only for a limited lapse of time, usually referred as 'Ehrenfest time' $t_E $ \cite{Ehren}. Above this time the modulus square of the wave packet spreads out of the classical trajectory and the correspondence principle appears to be invalid. In a region of unstable classical trajectories this time is expected \cite{Berman} to be not longer than a limiting time which increases as $ ln(S/\hbar) $ in the limit $ \hbar \rightarrow 0 $. Here $S$ is a typical classical action of the system. Therefore, for macroscopic system the correspondence principle appears to be untenable even for relatively short time. The classical limit seems to need a clarification. \par 
A second related problem in the classical limit is the so called "Schroedinger cat paradox" \cite{Weinberg}, which arise if the superposition principle is extended to macroscopic objects. This "paradox" suggests that the transition from quantum to classical regime is a real one rather than an approximate one, however precise it can be. The problem appears connected to the measurement problem, since the understanding of reduction of the wave function is clearly involved.

\vskip 0.3 cm
\par\noindent
{\it Null experiments.} 
\vskip 0.3 cm
\par 

In some cases the reduction process can occur without interaction. The typical experimental arrangement is the one proposed in ref. \cite{EV,V}. Let us consider a Mach-Zender interferometer, where the optical paths for the two branches
are tuned to have a destructive interference in one of the output port, so no photon can be detected there. If in one of the two branches an absorbing object is introduced, there is a 50\% probability that the photon is absorbed and 50\% probability that is not. In the first case still no photon is observed in the output port. In the second case the photon must have passed in the other branch, in which case the interference cannot be present and the photon can be detected at the port. The problem arises in the second case, since the photon wave packet has been reduced to the component in the branch which is free of object. This means that the reduction process has occurred without any interaction. The situation is clearly different from the usual measurement process, where the reduction takes place at the time of interaction with the  detector. 
An experimental realization of this proposal was performed in ref. \cite{Kwi} with a Michelson interferometer and a special device to handle single photon states. Asymmetric beam splitter was also considered, and the results are in agreement
with quantum mechanical predictions.
If we pretend that no physical process can develop without interaction, the null experiment just described suggests that the reduction process must involve some sort of 'hidden interaction', which however has not to involve
local 'hidden variables', since, as it is well known, they are not tenable \cite{Bell}. 

\vskip 0.3 cm
\par\noindent
{\it Spontaneous decay.} 
\vskip 0.3 cm
\par 

It can be a little surprising the inclusion of an elementary and well known phenomenon among the question raised by QM.
To exemplify the problem, let us consider an isolated atom that is in an excited state that can decay only to the ground state. If we follow the ordinary QM formalism, the initial state evolves by developing two components, one corresponding to the excited state and one corresponding to the ground state plus a photon. The strengths of the two components change with time, with the excited state component decreasing in time and the other increasing in time, keeping the overall normalization constant. If at a certain distance a perfect detector is present, the photon will be observed there and the atom will be reduced to the ground state. If $L$ is the distance of the detector, the detection will occur at the time $ t = L/c $, with $c$ the velocity of light (we are neglecting the size of the photon wave packet with respect to $L$ ).
If we repeat the procedure, the photon will be always detected at the time $t$ after the atom excitation. This is in contrast with observation, where an exponential decay law is clearly present, i.e. the probability that the photon is emitted before the time $t$ increases as $ 1 -  \exp(-t/\Gamma) $, where $ \Gamma $ fixes the decay rate. To accommodate this finding into QM, one assumes that the atom decay occurs randomly with probability density per unit of time equal to $ \Gamma $. Also in this case one is forced to introduce additional hypothesis, not included in the ordinary QM formalism.\par 
The decay process is an example of "quantum jumps", where the state of the system changes abruptly \cite{jump1,jump2,jump3}. It is usually assumed that the quantum jumps are instantaneous \cite{jump1,jump2,jump3}, and phenomenology looks compatible with this hypothesis. However, one has to distinguish between real quantum jumps, like photon emission, and general electron dynamics, which has indeed been probed within a time scale at the attosecond level \cite{atto1,atto2}.  
\par 
Reduction itself can be thought as a quantum jump, since it cannot be described within the QM formalism and its dynamics cannot be probed. 
\vskip 0.3 cm
\par\noindent
{\it Entanglement.}
\vskip 0.3 cm
\par 

Entanglement between two correlated particles, as illustrated by the experiment proposed by Einstein, Podosky and Rosen
\cite{EPR,Bohm}, can be considered an unavoidable  consequence of the wave function reduction produced by measurements. The correlation between the results of the measurements on the two distant particles is a consequence of the reduction and of the extension of the two particle wave function.  The reduction selects one of the two possible outputs for one of the two particles, while the extension of the wave function produces an effect at a distance, which is the puzzling aspect of entanglement. The non-locality emphasized by the violation of the Bell' s inequality is therefore intimately linked with the measurement problem.\par 
Entanglement requires an extended separate discussion, which will be given elsewhere.

\subsection{The method.}
In this subsection we sketch the method used to extend the mathematical formalism and physical contents of ordinary QM.
\par 
Since the pioneering work by Robinson \cite{Robinson1,Robinson2}, it has been established that it is possible to extend properly the standard analysis of real numbers into a nonstandard analysis which introduces additional numbers. The new nonstandard numbers include infinitesimals and unlimited numbers. The former are numbers smaller than any standard real numbers, the latter are numbers larger than any standard real numbers. The nonstandard analysis can be developed either within model theory
or within an axiomatic approach. The nonstandard sector introduces mathematical objects that do not exist in the standard sector. The method we follow is to give a physical meaning and a physical role to these objects, which necessarily
implies new physics. Generally speaking, there is no reason for forbidding the use of nonstandard analysis, and treat the nonstandard numbers on the same footing as the standard one, provided this is done consistently and carefully.\par 
A brief and schematic introduction to nonstandard analysis is provided in Appendix A. For a more extended exposition of nonstandard analysis one can consult textbooks on the subject \cite{Keisler,Goldblatt,Albeverio,Sousa,Diener,Nelson,Loeb,Machover,Benci}.\par 
Within this general framework, the so called hyperfinite real axis ${\rm \Pi}$  is introduced, following e.g. ref. \cite{Albeverio,Sousa}.
It consists of a hyper lattice along the real axis, where the lattice spacing is infinitesimal. This is done for both the three-dimensional coordinates and the time axis, but with a different order of the infinitesimals. These lattices are supposed to be the support of the physical phenomena. The space lattices have definitely not be confused with the discrete numerable lattice often introduced in the discrete geometry used to describe physical phenomena. Loosing speaking, the number of the hyperfinite lattice points is infinitely higher than the one of the standard real axis. In fact, if we consider the monads of \Rns restricted to the hyperfinite real axis \mdPi, each one contains an unlimited number of points of ${\rm \Pi}$.  
To take the limit to zero of the lattice spacing of a numerable representation of the real axis corresponds to the continuum limit and one gets just the standard QM in \Rs\ .
\par
This type of hyper lattice has been recently used in ref. \cite{Kass}, where the 'Probabilistic Bohmian Quantum Mechanics'
has been developed. We also use this lattice as the starting point of the model, but we follow a different scheme and a different conceptual framework.
\par
Each wave functions are supposed to have a standard part and a nonstandard part. The introduction of the nonstandard component is equivalent to consider the possibility of singularities in the wave function in addition to the standard smooth component of ordinary QM. 
From the wave equation, extended to the nonstandard sector, one can construct plain waves of nonstandard infinitesimal wavelengths and unlimited frequencies, as well as the nonstandard part of a generic state wave function. The unlimited frequencies are necessarily not related to energies, but are relevant for the evolution of the quantum state. Both the lattice spacing and the nonstandard plain waves are not observable quantities. However, following the procedures typical of ordinary statistical mechanics, we will show that they can be the basis for the appearance of a stochastic Ito process along the standard time-axis. This stochastic evolution implements the Schrodinger equation and it is able to trigger the reduction process in a measurement and entails the Born' s rule of QM. Within this scheme one finds that there must be a critical size of a physical object above which no quantum superposition among the states of the system is possible, which marks the transition from the quantum to the classical regime. This critical size fixes the boundary between the microscopic and the macroscopic worlds. Similarly, the null experiments, the spontaneous decay and the quantum jumps find coherent natural explanations.  
In the next sections the model will be built and proposed. Along the different steps for the construction of the model,
few assumptions will be formulated. Some can appear \emph{ad hoc}, introduced just to obtain the desired results. However one has to keep in mind that standard quantum mechanics developed on the basis of phenomenological observations. At the very beginning of Quantum Mechanics development the quantization of the orbits was devised by N. Bohr to reproduce the hydrogen atom spectrum. The introduction of the wave functions was suggested by the experiments on interference and diffraction.
The general canonical quantization was formulated in order to get the transition from classical to quantum mechanics.
All these hints finally converged into the formulation of Quantum Mechanics that we know today, but its
structure clearly reflects the apparently arbitrary assumptions that are at its origin. Since the model complete Quantum Mechanics, some additional assumptions are necessarily needed.

\section{Building the Model.}
\subsection{The nonstandard component.\label{sec:NScomp}}
In standard QM one cannot fix the position of a particle with arbitrarily accuracy, because of the uncertainty principle,
which then would imply infinite momenta. However, the physical space, which is the support of the wave function, is a real number manifold of suitable dimensions. 
We extend the manifold to the nonstandard sector, so that each one of the spacial and time coordinates are defined by the nonstandard real axis \Rns\, instead of the standard real axis \Rs\ . One can also introduce the set \Nns\ of unlimited integer numbers, which can be defined as integer numbers larger than any standard integer number. \par  
Following e.g. ref. \cite{Albeverio,Sousa,Kass}, we consider  in \Rns\ the so-called hyperfinite real line ${\rm \Pi}$.
Fixing an unlimited integer number $N$, the ratio
\beq
d \,=\, 1/N
\eeq   
\noindent
is an infinitesimal. Here in the numerator the number $1$ is intended to indicate some unit of length. We then consider the numbers
\beq
 x_j \,=\, \frac{j}{N} \,\equiv\, j\times d \ \ \  ;  \ \ \ j \,=\, -M, -M+1\, -M+2, ........ M-2, M-1, M  
\label{eq:points}\eeq 
\noindent 
with $ M \,=\, N^2 $. The $2M + 1$ numbers are equally spaced points in \Rns\tt, which include both infinitesimal and
unlimited numbers. The set of the standard parts of those numbers which are finite can be embedded in \Rs\tt.
The main assumption is that ${\rm \Pi}$ is the support of the wave function, i.e. each coordinate can take only the values in ${\rm \Pi}$. For the time axis we do the same, but the corresponding hyperfinite line ${\rm \Pi}_t$ is assumed to have an infinitesimal time step $ d_t $ of higher order than $ d $. This is a basic and essential assumption that has fundamental consequences on the structure of the model. 
\par The sizes of $ d $ and $ d_t $ do not fix any scale at the standard level, and it is required only that they are infinitesimals. At the nonstandard level of \Rns\ ${\rm \Pi}$ 
is not translationally invariant nor isotropic, but at the standard level of \Rs\ the invariances are maintained, and it is so for the invariance for time translation. \par  
For proceeding further we need the wave equation for the wave function $ W(x) $. For sake of simplicity let us consider the case of a massless and spinless particle, so that the equation is of the Klein-Gordon type. More general cases need only minor modifications. Let us consider first a free particle. On ${\rm \Pi}$ in one dimension the equation can be written
\beq
 \frac{D^2 W_n(t)}{D t^2} \,=\, c^2 \frac{1}{d^2} (W_{j+1} - 2 W_{j} + W_{j-1}) 
\label{eq:w1}
\eeq  
\par\noindent
where $W_j = W(x_j)$ and $c$ the speed of light. In the following we put $c \,=\, 1$. The index $j$ runs from $-M +1$ to $M - 1$, for a total of $2M - 1$ equations. The values of $W_{-M}$ and $W_M$ must be fixed by the boundary conditions. The right hand side of Eq. (\ref{eq:w1}) is just the second derivative in discrete form, but with an infinitesimal step size. At the left side one has also the non standard second derivative in time on ${\rm \Pi}_t$, defined analogously. 
The tridiagonal band matrix at the right side can be diagonalized to get both eignvectors and eigevalues. 
 As shown in Appendix B, the eigenvectors of the matrix are
\beq
\phi_k(x_j) \,=\, \frac{1}{\sqrt{2M+1}} \exp(\frac{\imath k j \pi}{M}) \ \ \ ; \ \ \ k \,=\, 1, 2 \cdots\cdots M 
\label{eq:ev}
\eeq 
\noindent
with eigenfrequencies 
\beq
\omega(k) \,=\, \pm \frac{2}{d} \sin(\frac{k\pi}{2M})
\label{eq:ef}
\eeq
The eigenvectors are taken orthonormal according to the scalar product, 
\beq
 (u,v) \,=\, \sum_j\, \overline{u(x_j)} v(x_j)
\label{eq:sp} 
\eeq
\noindent
for two generic vectors $u,v$. The solutions of the wave equation are then
\beq
\Psi_k(t,x_j) \,=\, \phi_k(x_j) \exp(\imath \omega(k) t)
\label{eq:solwe}\eeq
\noindent as it follows from the calculation of the nonstandard time derivative
\beq
\frac{D}{D t} \exp(\imath \omega(k) t) = \frac{1}{d_t} \exp(\imath \omega(k) t)(\exp(\imath \omega(k) d_t) -1 )
 = \imath \omega(k) \exp(\imath \omega(k) t) 
\label{eq:nsder}\eeq
\noindent and so for the second derivative. In fact, since $ \omega(k) $ is of the order at most of $ 1/d $, and
$ d_t/d $ is an infinitesimal, $ \omega(k) d_t $ is also infinitesimal.  
\par 
If we introduce the wave vector $ p = k \pi/N $, the spatial part of the eigenfunctions can be written in the usual way
\beq
\phi_p(x_j) \,=\, \frac{1}{\sqrt{2M+1}} \exp(\imath p x_j) 
\label{eq:wavec}\eeq
\par\noindent and accordingly the eigenvalues $\pm \omega(p) = \frac{2}{d} \sin(\frac{p}{2N}) $.
\par 
One can notice two distinct regions in the spectrum. If the argument of the sine function is infinitesimal, one can substitute the sine function with its argument, and the frequency is linear in the wave vector. This region includes the standard spectrum a well as an extension to the nonstandard sector of unlimited frequencies and momenta. This region include all the wave vectors of the order of $ N^{1-\eta} $, with $ 0 < \eta < 1 $.
This region is just a proper extension of the linear standard spectrum. 
\par The second region is the one characterized by $k$-values in Eq. (\ref{eq:ef}) of the type $ k = [ q N^2 ]$, with $ q $ a finite number smaller than $ 1 $ and the square parenthesis indicates the integer part. In the following the integer part is implied when necessary. In this region the frequency is a nonlinear function of the wave vector. \par 
The first region can be considered trivial and we assume that only the second region is relevant in constructing the nonstandard part of a generic wave function.\par 
For consistency, the nonstandard frequencies and wave vectors cannot be related to energy and momentum, respectively, as in the standard sector, since they are unlimited quantities. However, they are related to period and wavelength in the usual way. \par 
For a generic localized single particle wave function, the nonstandard part of the wave equation has a the general solution
(Appendix B)
\beq
\Psi^{NS}(t,x_j) \,=\, \sum_l \exp(\imath p_l x_j) \cos(\omega(p_l)t) ^*\Psi^S(t,x_j)
\label{eq:ansatz}\eeq
\noindent where  $\Psi^{NS}$ is the nonstandard part of the wave function, $^*\Psi^S$ is the canonical extension of the standard part of the wave function, and $x_j$ are the points of Eq. (\ref{eq:points}). The use of the cosine function corresponds to the assumed localized wave function, which requires the restriction to standing waves. 
Notice that the nonstandard part must be necessarily time dependent. In fact the energy eigenvalues must be finite and therefore they  pertain to the standard part of the wave function.
\par
The set of wave vectors $p_l$ belong to the nonlinear part of the spectrum. There is an ample freedom for the choice of the set of wave vectors, leaving all the results unchanged. Our specific choice is introduced here only to serve as a proof of the possibility for the model to work. We take an array of wave vectors (positive part)
\beq
p_l \,=\, p_0 \,+\, l \Delta p, \ \ \ \ \ \ l = 0, 1, 2\cdots\cdots l_0
\label{eq:array}\eeq
\noindent
with 
\beq
p_0 \,=\, q_0 \pi N, \ \ \ \ \ \  p_1 \,=\, p_0 \,+\, l_0 \Delta p \,=\, q_1 \pi N
\label{eq:range1}\eeq
\noindent where $0 < q_0 < q_1 < 1$ are given finite numbers. Notice that $\pi N$ is the maximum value of the wave vectors. The step $\Delta p$ is assumed unlimited, as well as $l_0$.
This is possible, since the range of wave vectors $(q_1 - q_0) \pi N $ is unlimited. To be specific, let us assume that 
$\Delta p \sim \pi N^\eta $, with $0 < \eta < 1$, which is unlimited. Then the number $N_p$ of $p$-values can be estimated as $N_p \sim (q_1 - q_0)N^{1-\eta}$, which is also unlimited. All that corresponds to a suitable array of $k$-values. 
Accordingly, the values of the frequencies span the unlimited interval  $\omega_0 <  \omega < \omega_1 $, with, for massless particles,
\beq
\omega_0 = \frac{2}{d} \sin(\frac{p_0}{2N})\ \ \ \ \ \ \ \ ; \ \ \ \ \ \ \ \  \omega_1 = \frac{2}{d} \sin(\frac{p_1}{2N})
\label{eq:spom}\eeq
\noindent
These expressions are still valid for massive particles since the mass term is infinitesimal with respect to the nonstandard frequencies.
\par 
The overlap between the nonstandard parts of two state wave functions can be defined as
\beq
< \Psi_1^{NS} | \Psi_2^{NS} > \,=\, d\sum_j\, \overline{\Psi_1^{NS}(x_j)} \Psi_2^{NS}(x_j)
\label{eq:nssp}\eeq
\noindent It is ready verified that this definition applied to the standard components merges into the usual scalar product between two standard wave functions. In Appendix B it is shown that this definition, applied to wave functions of
Eq. (\ref{eq:ansatz}), gives
\beq
< \Psi_1^{NS} | \Psi_2^{NS} > \,=\, < \Psi_1^S | \Psi_2^S > \sum_l \cos(\omega(k_l)t)^2
\label{eq:nsnorm}\eeq  
\noindent
A straightforward consequence of Eq.(\ref{eq:nsnorm}) is that the modulus square of the nonstandard component cannot be interpreted as a probability distribution. In fact, as shown in the sequel, the Born' s rule can be deduced by the model, so its inclusion here would be a tautology.\par  
Notice that the overlap of Eq. (\ref{eq:nssp}) between a standard and a nonstandard components vanishes, since 
they belong to two orthogonal space of wave vectors. \par 
A physical state of a given system will be represented by a wave function that can be written as a linear combinations of a basis functions with a nonstandard component (\ref{eq:ansatz}). 
The coefficients of the expansion will be determined only by the standard components, since the nonstandard components are then automatically defined. In other words if one expands
\beq
\Psi \,=\, \sum_l \, C_l \, \psi_l 
\label{eq:expan}
\eeq 
\noindent with $ \psi_l$ a basis state, then 
\beq
C_l \,=\, < \psi_l^S \vert \Psi^S >
\label{eq:coef}
\eeq
\par
The expression of Eq. (\ref{eq:ansatz}) can be generalized to many-body systems. Generally speaking one can apply the same expression for the coordinate $x_j$ of each particle $j$. However what has to be considered is the degrees of freedom that
must be used to describe the system. In particular one can consider a collective excitation of a many-body system, in which case the collective degrees of freedom is the only relevant variable. As an example let us consider a localized acoustic wave \cite{qtest}, where the system can be described by the collective displacement variable $R$ and the corresponding standard  wave function $W(R,t)$ for a given oscillatory mode. The displacement of each atom will be function of $R$, so that the system ha only one effective degrees of freedom. The nonstandard component of the state wave function will be then
\beq
W^{NS}(R,t) \,=\, \sum_l \exp(\imath p_l R) \cos(\omega(p_l)t) ^*W(R,t)
\label{eq:acoustic}\eeq 
\noindent 
where the canonical extended wave function $^*W(R,t)$ can be taken on the hyperfinite lattice sites.
This expression is justified as before since $W(R,t)$ is expected to satisfy a wave equation.
As another example one can consider the excited states of a superfluid at low energy \cite{Leggett1,Leggett2}. The system can be then described simply by the local superfluid amplitude $\psi(r,t)$, with only one degrees of freedom. In particular the current flow
is embodied in this single variable wave function, with a nonstandard part as in Eq. (\ref{eq:acoustic}).
\par
The normalization of Eq. (\ref{eq:nsnorm}) requires some physical justifications. 
First of all the result of Eq. (\ref{eq:nsnorm}) can be generalized to many particle states.
To this respect one must notice that the overall wave function of two independent systems is the product of the wave functions of each system, but the standard and nonstandard parts are multiplied separately. In other world, no cross product of a standard and nonstandard part must be present. This is required by the different nature of the two components, since a cross term would have an ambiguous meaning. 
As a consequence the strength of the nonstandard component increases with the number of degrees of freedom. Furthermore, as shown in Appendix C, each degrees of freedom gives to the nonstandard interaction matrix elements a factor
proportional to the number of p-values, which is supposed to be unlimited. It can surely appear that this normalization is an \emph{ad hoc} postulate. However, it turns out that there is not so much other choices for the normalization if one has to obtain quantum jumps as a possible physical process of a quantum system. In fact not only
quantum jumps can then develop during the evolution of a system, but also they must be necessarily instantaneous.
All that appears to be in accordance with phenomenology. \par 
No other choice of the normalization could achieve instantaneous quantum jumps. In particular a finite value of the normalization would result in a vanishing small interaction matrix elements for any size of the system, with no effect on the evolution of the quantum state. These considerations suggest that the nonstandard part of the wave function has an ontological character, since its intensity increases exponentially with the number of degrees of freedom, i.e. it shares 
some properties with a physical "field". Of course this is only a question of interpretation, with no consequence on the formalism.
\par
On a more mathematical side, the result of Eq. (\ref{eq:nsnorm}) shows that with a finite normalization the functional space
of the nonstandard parts of the wave functions would be just a replica of the one of the standard parts. A genuine extension of the functional space requires the presence of singularities in the nonstandard part of the wave function, as shown explicity in Appendix B. More technically, the additional functional space must lye outside the nonstandard hull \cite{Hurd} of the nonstandard functional space. 
\par
All that is apparent once 
the separation between standard and nonstandard components of the wave function is extended to the density matrix $ \rho(t,x_j,x_{j'}) $. This means that the nonstandard component 
$ \rho_{NS}(t,x_j,x_{j'})$ must contain only nonstandard frequencies, which implies the expression
\beq
 \rho_{NS}(t,x_j,x_{j'}) \,=\, \sum_{r\neq r'} \cos(\omega_r t)\cos(\omega_{r'} t) \exp(\imath ( (p_r - p_{r'}) x_j ) ^*
\rho_S(t,x_j,x_{j'})
\label{eq:rons}
\eeq 
\noindent where $ ^*\rho_S $ is the canonical extension of the standard density matrix. Of course, for consistency, the diagonal term
$x_j = x_{j'}$, is time dependent and cannot be related to the density of a stationary or non-stationary state. \par  
The nonstandard density matrix can be defined along the same lines for a many particle system. The interaction matrix elements between the nonstandard components of two states $ l, m $ can then be calculated, as described in Appendix C.
\subsection{The fundamental stochastic process.\label{sec:fpro}}
In this section we introduce, on the basis of physical considerations, the fundamental stochastic process of the matrix elements which describes the interaction between the nonstandard components of physical states. 
We work in the interaction representation with respect to the standard interaction, so that 
the time evolution is determined by the interaction matrix elements between the nonstandard parts of the wave functions. \par
 The finite part of \Rns{ } is the disjoint union of all the monads in \Rns, and so  ${\rm \Pi}$ is the disjoint union
of all \mdPi. All points of a monad have the same standard part $t$, which acts as a label of the monad.
In each \mdPi\!\!(t) we consider the
point $r_0(t)$ that is just to the right of $t$ such that $0 \leq  | r_0  \,-\, t | \leq d_t/2$ and the set 
$ I_{n_0}(t) $ of points
\beq
r(t,j) \,=\, r_0(t) \,+\, j d_t \ \ \ \ \ \ \ \ \ -n_0 \, \leq j \, \leq n_0
\label{eq:pmod}
\eeq     
\noindent where $ n_0 $ is an unlimited integer number such that $a_0 =  n_0 d_t \in\ $\mdPi, to be used as a cutoff. The physical results will not depend on its particular value. The need of such a cutoff stems on the fact that one cannot identify the boundaries of a given monad. We call 
${\rm \Pi}_t^{(n_0)} $ the hyperfinite real axis restricted according to the cutoff.
\par 
The matrix element  $ \xi_{lm}(t) $ of the interaction between the nonstandard components of two physical states $ l, m $ (see Appendix C) can be written
\beq
\xi_{lm}(t) \,=\, {\cal I}^{lm} \sum_p^{N_\Omega} \exp(\imath \phi_p^{lm}) \exp(\imath \Omega_p t)
\label{eq:inme}\eeq
\noindent
where $ {\cal I}^{lm} $ is a positive definite quantity, $ \Omega_p $ are an unlimited set of $ N_\Omega $ nonstandard frequencies and $ \phi_p^{lm} $ a phase that depends on the frequency but not on time. The summation over the frequencies is rapidly and irregularly oscillating within a \mdPi. For the development of the model it is convenient 
to introduce the reduced matrix element $ X_{lm} $
\beq
X_{lm} \,=\, \xi_{lm}/{\cal I}^{lm}\sqrt{a_0N_\Omega} 
\label{eq:rme}\eeq
\noindent where the normalization was chosen for later purpose.
\par 
The main part of the construction of the stochastic process is the transition from the
time flow in \Rns to the one in \Rs. 
All the values of the time in a monad$_{\rm \Pi}$ have the same standard part.
We assume that the flow of time of the quantum processes occur in \Rs. This is because infinitesimal
time intervals have no measure. In other words in the model there is a sub-quantum evolution 
that is below the quantum one and entails the stochastic part of the evolution in \Rs.
 The transition is done in two steps. 
\par\noindent 
1. In a given interval of time, e.g. in an experiment, a representative point 
   in each \mdPi is selected. In the set of all runs the points of a \mdPi are chosen with equal         
    frequency. In other words each \mdPi is the support of a stochastic process with
   uniform probability distribution. Stochasticity in this case can be considered
   a consequence of the uncertainty to fix the time value within a \mdPi. This is a procedure
   quite similar as in Statistical Physics, where the lack of knowledge of the microscopical
   details of the system configurations is translated into a probability distribution of the
   microscopical configurations \cite{Balian}. In particular for the microcanonical ensemble 
   one indeed takes a uniform distribution over the microscopic configurations compatible 
   with a given energy.  
\par\noindent 
2. In each run of the experiment the value of the reduced matrix elements in a \mdPi are taken constant and equal 
   to the one at the representative point. This can be viewed as a ''coarse graining" procedure, also mediated
   from non-equilibrium statistical mechanics \cite{Balian}, in which the phase space is divided into a set of small cells 
   (or projection operators). Inside each cell the distribution is considered uniform and,
   accordingly, the statistical density matrix is taken proportional to the unit matrix.   
   This allows the transition from \Rns to \Rs. In this way the stochastic process on the points
   of each \mdPi induces the stochastic evolution of the matrix elements.
\par 
On the basis of these physical considerations we introduce the following stochastic process      
 of the reduced matrix element  $ X_{lm}(t) $ of the interaction between the nonstandard components of two physical states $ l, m $ (see Appendix C)
\begin{displaymath}
\left\{ \begin{array}{ll}
X_{lm}(r(t,j))&\,=\, X_{lm}(r(t,j_t)) \ \ \ {\rm for\,\, all\,\, } r(t,j)  \\  
\ &\ \\
X_{lm}(r'))&\,=\, 0 \ \ \ \ \ {\rm for\,\, all\,\, others} \ \ r' \in  {\rm monad}_\Pi  \\ 
\end{array}\right .
\label{eq:matst}
\end{displaymath}
\par\noindent The index $ j_t $, and the corresponding point $ r(t,j_t) $, are treated as random variables for each monad with uniform probability, which fixes an independent stochastic process for each \mdPi. \par 
This stochastic process is a constitutive element of the model, which is fixed by
the physical considerations above. It can be viewed as originating by a statistical treatment of the evolution in {\Rns}.
When going back to the matrix element $ \xi_{lm} $
for consistency with the coarse graining procedure we will replace the factor $ {\cal I}^{lm} $ by its average over a \mdPi.
\par\noindent 
%
We have seen that the single particle frequencies span an unlimited interval with a minimum value $\omega_0$ and a maximum one $\omega_1$. Furthermore the corresponding wave numbers increase with an unlimited step $\Delta p$, so the differences between two frequencies are unlimited.
\par 
The frequencies appearing in the nonstandard matrix elements are combinations of single particle frequencies.
In Appendix C it is shown that also these frequencies are unlimited and the differences between two of them are larger than an unlimited minimal value $\Delta \Omega $, provided $\Delta p \sim N^\eta$, with $\frac{1}{2} < \eta < 1$.
\par 
One can assume that the range of the frequencies is such that 
\beq
\Omega_0 \,\leq \Omega_p \,\leq \Omega_1
\label{eq:range}\eeq
\noindent where $ \Omega_1 \sim d^{-1} $ and $ \Omega_0 \sim d^{-\delta},\ 0 < \delta < 1$. Here the symbol $ \sim $
indicates equality of the order of two unlimited numbers, i.e. their ratio is a finite number.
Then we first proof the following general lemma 
\par\noindent
LEMMA
\par 
If the (unlimited) minimal spacing $ \Delta\Omega $ between the frequencies $ \Omega_p $ is such that 
$ \Delta\Omega > \Delta\Omega_0 \sim d^{-\alpha} $, for some $ 0 < \alpha < 1$, then the stochastic process $ X_{lm}(u) $, where $ u $ is a generic point on $ {\rm \Pi}_t^{(n_0)}  $, is the nonstandard representation of a white noise process.
\par\noindent
The hypothesis of the lemma are actually satisfied in the model, as shown in Appendix C.\par\noindent
Proof
\par\noindent
The process $ X_{lm}(u) $ depends on the pair $ l,m $, but for simplicity we omit the indexes, keeping in mind that the processes for different pairs are independent.\par 
The frequencies $ \Omega_p $ are non-zero linear combinations of the single particle frequencies $ \omega(k) $,
which are unlimited in numbers. Therefore the number $ N_\Omega $ is also unlimited. According to the hypothesis of the lemma, one can estimate an upper limit for $ N_\Omega $ 
\beq
N_\Omega \,<\, ( \Omega_0 - \Omega_1 )/d^{-\alpha} \,=\, d^{-1+\alpha} \,-\, d^{-\delta+\alpha} \,<\, d^{-1+\alpha}  
\label{eq:upper}\eeq 
\par 
\noindent
One can assume $N_\Omega \,\sim\, d^{-\beta} $ for some $ 0 < \beta < 1 $.
Let us  consider the ensemble average of the process $ X(t) $ within a given \mdPi(t). For each time $r(t,j)$ of 
Eq.(\ref{eq:pmod}) the matrix element can take the values
\beq
X(r(t,j_t)) \ \ \ ; \ \ \ j_t \,=\, -n_0, -n_0+1, \cdots\cdots n_0-1, n_0 
\label{eq:mean}
\eeq     
\noindent
with uniform probability. The mean value of $ X $ is then
\beq
 < X > \,=\, \frac{1}{2n_0}\sum_{j=-n_0}^{n_0} X(r(t,j)) \\
\label{eq:mean1}
\eeq
\noindent Taking the expression (\ref{eq:rme})  for $ X $ the summation in Eq. (\ref{eq:mean1}) reduces to a geometric series. The final results can be written
\beq
< X > \,=\, \frac{1}{n_0}  \sum_{p=1}^{N_\Omega}\frac{\sin(\Omega_p d_t(n_0 + 1/2))}{\sin(\Omega_p d_t/2)} \exp(\imath \Omega_p r_0(t))/\sqrt{a_0 N_\Omega}  
\label{eq:mean2}
\eeq
\noindent where the sum is restricted to the positive frequency only, in ascending order. The product $ \Omega_p d_t $ is still an infinitesimal, because of the higher order of the infinitesimal step $ d_t  $, and therefore the sine function in the denominator can be substituted by its argument. The denominator is then $ n_0 \Omega_p d_t $. Since $ n_0 $ is a cutoff, it can be chosen large enough to make unlimited the denominator, but still keeping $ n_0 d_t $ infinitesimal. 
 An upper limit of the absolute value of $ < X > $ can be easily estimated
\beq
 | < X > | \,<\, \frac{1}{\sqrt{a_0 N_\Omega}}\sum_r \frac{1}{\Omega_r a_0}
\label{eq:upmean}\eeq 
\noindent
We now show that $ | < X > | $ is infinitesimal. In fact
\begin{eqnarray}
| < X > | &\,\leq\,& \frac{1}{\sqrt{a_0 N_\Omega}} \frac{1}{a_0} \big[ \frac{1}{\Omega_1} + \sum_{r > 1} \frac{1}{\Omega_r} \big] \nonumber\\ 
\ &\ &  \nonumber\\ 
\ &\,\leq\,& \frac{1}{\sqrt{a_0 N_\Omega}} \frac{1}{a_0\Omega_1} \big[ 1 + \sum_{r \geq 1}\frac{1}{1 + r\Delta\Omega/\Omega_1 } \big] \nonumber\\ 
\ &\ & \nonumber\\
\ &\,\leq\,& \frac{1}{a_0^{3/2} N_\Omega^{1/2}} \big[ 1 + \sum_{r \geq 1}\frac{1}{r\Delta\Omega/\Omega_1 } \big] \nonumber\\ 
\ &\ & \nonumber\\ 
\ &\,=\,& \frac{1}{a_0^{3/2} N_\Omega^{1/2}} \big[ \frac{1}{\Omega_1} + \frac{1}{\Delta\Omega} \ln (N_\Omega) \big] 
\label{eq:mean3}\end{eqnarray}
\par\noindent
As discussed above, one can take $ a_0 \sim d^{-\gamma} $, with $ 0 < \gamma << 1 $. Then 
\beq
a_0^{3/2} N_\Omega^{1/2} \,\sim\, d^{3\gamma/2 - \beta/2}
\label{eq:expon}\eeq
\noindent
In the limit of a large cutoff for $ n_0 $, within the allowed range, $ \gamma < \beta/3 $ and this product is unlimited. 
Then the final expression of Eq.(\ref{eq:mean3}) is infinitesimal, since the logarithmic divergence is of lower order than any power divergence.  
\par 
We now turn to the correlation function $ \chi(u,u') = < \overline{X(u)} X(u') > $, where the average is the ensemble average and the horizontal line indicates complex conjugate. First of all,
if $ u $ and $ u' $ do not belong to the same monad the correlation function is infinitesimal. In fact the process in each monad is independent on the others, so that the correlation function reduces to the product of two averages, i.e. 
\beq
 < \overline{X(u)} X(u') >  \,=\, < \overline{X(u)} > < X(u') > ,
\eeq 
\noindent  and each factor of the product is infinitesimal, as shown above. \par 
Let us then assume $ u $ and $ u' $ in the same monad. Because the function $ X(u) $ is constant inside each monad,
the correlation function will not dependent on the values of $ u $ and $ u' $, and one can take $ u \,=\, u' $,
and the correlation function reduces to the average of $ | X(u) |^2 $, which is constant for $u$ inside the considered monad, 
\begin{eqnarray}
< \overline{X(u))} X(u') >  &\,=\,& < | X(u) |^2 > \,=\, \frac{1}{a_0 N_\Omega} < | \sum_r \exp( \imath \Omega_r u ) |^2 > \nonumber\\
\ &\ & \nonumber\\
 &\,=\,& \frac{1}{a_0 N_\Omega}  < \sum_{r r'}  \exp( \imath ( (\Omega_r - \Omega_{r'})u ) > \nonumber\\
\ &\ & \nonumber\\
 &\,=\,& \frac{1}{a_0 N_\Omega}  \sum_{r r'} \frac{1}{2n_0} \sum_{j_t} \exp( \imath ( (\Omega_r - \Omega_{r'})
 j_t d_t )  
\label{eq:eqcorr}\end{eqnarray}  
\noindent where the summation over $ j_t $ is in agreement with Eq. (\ref{eq:mean}). Following now the same procedure leading to the infinitesimal character of $ <  X(u) > $, see Eq. (\ref{eq:mean3}), one finds that the sum $ S $ of the terms with $ r \,\neq\, r' $ is infinitesimal
\beq
S \,<\,  \frac{2}{a_0^2\Delta \Omega N_\Omega} \ln(N_\Omega) \,\sim\, {\rm monad(0)}
\label{eq:nond}\eeq
\noindent
The terms with $ r \,=\, r' $ sum up to $ N_\Omega $ and then finally the correlation function in this case is given by
\beq
< \overline{X(u))} X(u') > \,=\, \frac{1}{2 a_0} \ \ ; \ \ \ \ \ \  u , u'  \in \ I_{n_0}(t)
\label{eq:eqcorr1}
\eeq   
\noindent
for some standard time $ t $.
This correlation function is the representation on $ {\rm \Pi}_t^{(n_0)}  $ of a Dirac delta function, i.e. it is a so called pre-delta function \cite{Hoskins}. In fact, let us consider a smooth finite function $ F(t) $ on \Rs, and its canonical extension $ F^*(u) $ to $ {\rm \Pi}_t^{(n_0)}  $. Then the following equations are valid for a generic point $ v \in I_{n_0}(t)$
\begin{eqnarray}
 \ st\big[\sum_j d_t F^*(j d_t) \chi(j d_t,v) \big] &=& st\big[\sum_{j=-n_0}^{n0} d_t F^*(j d_t + r_0(t)) 
 \chi(j d_t + r_0(t),v) \big] \nonumber\\
\ &\, & \nonumber\\
 \ &=&  st(F^*(v)) = F(t) = \int du \delta(u-t) F(u) 
\label{eq:predel}\end{eqnarray}  
\noindent where we used the smoothness of $ F $. The initial summation is over the finite part of $ {\rm \Pi}_t $, i.e. to all the integers $ j $ such that 
$ j d_t $ is finite.
\par\noindent
Finally let us consider the spectral function of the process. For any number $ \Lambda $, we can consider
the function $ \exp(\imath \Lambda u) $ on the finite part of $ \Pi_t^{(n_0)} $ and the corresponding Fourier transform
\beq
d_t \sum_j \exp(\imath \Lambda j d_t) <\overline{X(j d_t)} X(v) > \,=\, 1
\eeq
\noindent for a generic point $ v $. This shows that the spectral function of the process is a constant, which defines the white noise. 
\par\noindent
This complete the proof.
\par\noindent  
We can now consider the question of the dependence on the cutoff $ n_0 $. Let us take a finite interval of time 
$ (u,v) $ on $ {\rm \Pi}_t  $ and its length
\beq
 v - u \,=\, \sum_{j=1}^n u \,+\, j d_t
\label{eq:length}\eeq 
\noindent where $ n = (v - u)/d_t $. If the summation in (\ref{eq:length}) is restricted to the points of 
$ {\rm \Pi}_t^{(n_0)}  $, the results will be a lower limit of the length. However the summation is a monotonically increasing function of $ n_0 $, with an infinitesimal step between $ n_0 $ and $ n_0 + 1 $. For large enough $ n_0 $, but still keeping $ n_0 d_t $ infinitesimal, $ {\rm \Pi}_t^{(n_0)}  $ must    
merge into $ {\rm \Pi}_t $. Therefore there must be a range of values of $ n_0 $ where the calculated length of the interval is infinitesimally close to $ v - u $. Even if we cannot estimate this range, this is the asymptotic region of the cutoff.
The structure of the reduction dynamics turns out to be cutoff independent. In the sequel  therefore we will not distinguish between $ {\rm \Pi}_t^{(n_0)} $ and $ {\rm \Pi}_t $. 
\par\noindent 
Associated with the white noise on $ {\rm \Pi}_t $, constructed above, one can introduce the stochastic process 
\beq
B(u) \,=\, \sum_{j=0}^{n} X(j d_t) d_t
\eeq
\noindent where $ u \,=\, n d_t $ and $ u \,=\, 0 $ is just a reference time.
\par\noindent 
THEOREM
 The process $ B(u) $ is the nonstandard representation of a Brownian process.
\par\noindent Proof
\par\noindent 
Let us consider the correlation function 
\beq
< \overline{B(u)} B(u') > \,=\, d_t^2\, \sum_{ll'} < \overline{X(l d_t)} X(l' d_t) > 
\eeq
\noindent On the basis of the conclusions of the Lemma, we can distinguish three cases
\par\noindent
i) \,\, st($u'$) $>$ st($u$)
\beq
< \overline{B(u)} B(u') > \,=\, d_t\, \sum_l \,1 \,=\, n d_t \,=\, u \,=\, st(u) \,+\, \eta_1
\eeq
\noindent with $ \eta_1 $ an infinitesimal.
\par\noindent
ii) \, st($u'$) $<$ st($u$)
\beq
< \overline{B(u)} B(u') > \,=\, d_t\, \sum_{l'} \,1 \,=\, n' d_t \,=\, u' \,=\, st(u') \,+\, \eta_2
\eeq
\noindent with $ \eta_2 $ an infinitesimal.
\par\noindent
iii)  st($u'$) \,=\, st($u$)
\beq
< B(u) B(u') > \,=\, st(u) \,+\, \eta_3 
\eeq
\noindent with $ \eta_3 $ an infinitesimal. It follows that, for $ u, u' $ finite
\beq
st( < \overline{B(u)} B(u') > ) \,=\, st(u)\wedge st(u')
\label{eq:wedge}
\eeq
\noindent where the symbol $ \wedge $ indicates the smaller between the two numbers. 
If we now consider, for each real standard time $ t $ of \Rs\ ,
the point $ u_t $ of \mdPi(t) just to the right of $ t $, as in Eq. (\ref{eq:pmod}), then one gets a stochastic
process $ {\cal B}(t) \,=\, st(B(u_t) $ on \Rs, with
\beq
 < \overline{{\cal B}(t)} {\cal B}(t') > \,=\, t\wedge t'
\label{eq:stochR}
\eeq    
\noindent Putting 
\beq
d{\cal B}(t) \,=\, {\cal B}(t+dt) \,-\, {\cal B}(t)
\label{eq:db}
\eeq
\noindent from Eq. (\ref{eq:stochR}) one gets 
\beq
< d\overline{{\cal B}(t)} d{\cal B}(t) > \,=\,  dt
\label{eq:var}
\eeq 
\noindent which is one of the constitutive equations of a Brownian motion and the only property of the stochastic process that will be used in the sequel. A similar construction can be found in ref. \cite{Keis_stoch} and in \cite{Albeverio}, Chapter 3.
Because the stochastic processes in different \mdPi are independent, we have also
\beq
< d\overline{{\cal B}(t)} > \,=\, 0
\label{eq:av0}\eeq 
\par 
The full interaction matrix elements include some factor $ F_{lm} $, to be made explicit later. Taking into account
the independence of the processes and restoring the index $ (l,m) $, Eq. (\ref{eq:var}) has to be replaced by
\beq
< d\overline{{\cal B}_{lm}(t)} d{\cal B}_{l'm'}(t) > \,=\, \delta_{ll'} \delta_{mm'} F_{lm}^2 dt
\label{eq:finalB}\eeq
\par 
We have implicitly assumed a pairwise interaction between the components.
\subsection{Stochastic and non-stochastic evolution. \label{sec:stoc}}
To be specific, let us consider the detection of the position of a particle. The detector can be e.g. a scintillator, that we will schematize with a set of identical molecules that can be excited by the particle interaction to a single excited state. The initial state of the particle will be a wave packet $ \Psi $, of much wider size than the resolution of the scintillator, moving toward the detector.  
One can expand the wave packet in a set of states, in particular a set of more localized wave packets $ \Psi_l $
\beq
 \vert \Psi > \,=\, \sum_l \vert \Psi_l > C_l     
\label{eq:inexp}
\eeq
\noindent where $ C_{l} $ is the overlap between the standard parts 
\beq
C_{l} \, =\, < \Psi_l^S \vert \Psi^S >
\label{eq:over}
\eeq 
\noindent 
For a single localized wave packet a definite time dependent state will be evolving. For the superposition of Eq. (\ref{eq:inexp}), if the evolution is linear and coherent, the time dependent state will be the linear combination of each evolved component. The nonstandard interaction between these components produces a stochastic evolution beyond the standard one. In this specific example the state that evolves in time corresponds to the excitation of a definite set of molecules of the detector. 
%
One can use the interaction representation to include implicitly the dynamics produced by the standard part only, i.e. one can introduce the evolution operator $ U_S(t)$ determined by the standard interaction only and the amplitudes of the overlap between the state $\Psi$  with the evolved basis states
\beq
C_{l}(t) \,=\, < \Psi_l \vert U_S^\dag(t) \vert \Psi >
\label{eq:amint}
\eeq    
\noindent The remaining evolution is determined by the nonstandard interaction. In view of the remarks above, one can write
\beq
\imath d C_l(t) \,=\, \sum_m d {\cal B}_{lm} C_m(t) 
\label{eq:stochev}
\eeq
\noindent where the stochastic process $ d{\cal B}_{lm}$, see Eq. (\ref{eq:finalB}), is directly related to the matrix elements of the nonstandard interaction between the nonstandard components (in the interaction representation). However, in view of Eq. (\ref{eq:var}), the square of the stochastic component can give contribution proportional to the differential $dt$, and, in agreement with Ito representation of stochastic process \cite{stoch}, the evolution equation we are looking for must be written in general as
\beq
\imath d C_l(t) \,=\, \sum_m d {\cal B}_{lm} C_m(t) + \imath\sum_m D_{lm} C_m dt
\label{eq:rightone}
\eeq 
\noindent where the matrix $ D_{lm} $ has to be determined on the basis of physical arguments. 
Stochastic evolution of the amplitudes have been considered in the literature \cite{Pearl1,Pearl2,Gsin1,Gsin2}. Here we are proposing the physical origin of the stochastic part of the dynamics. In order to specify the matrix $ D_{lm} $ on physical grounds we demand that the equation (\ref{eq:rightone}) conserves the normalization of the standard part of the state wave function in the expansion of Eq. (\ref{eq:inexp}) in each stochastic trajectory, as in standard QM. This is in conformity with the notion itself of quantum state. Without this constraint in fact the normalization could blow up or go to zero.
Let us consider the conjugate of Eq. (\ref{eq:rightone})
\beq
\imath d \overline{C}_l(t) \,=\, - \sum_m d \overline{{\cal B}_{lm}} \overline{C}(t) + \imath\sum_m \overline{D}_{lm} 
\overline{C}_m dt
\label{eq:rightonec}
\eeq 
\noindent and apply the Ito algebra for the calculation of the stochastic differential of a composite variable. One gets
for the variation of the modulus square of each component
\beq
\begin{array}{ll}
d(\overline{C}_l C_l) &\,=\, \overline{C}_l dC_l \,+\, C_l d\overline{C}_l \,+\, d\overline{C}_l dC_l  \nonumber \\
\ &\, \\ \nonumber
\ &\,=\, \overline{C}_l  ( - \imath \sum_m d {\cal B}_{lm} C_m \,+\, \sum_m D_{lm} C_m dt ) \\ \nonumber
\ &\, \nonumber \\
\ &\,\ +\, C_l ( \imath \sum_m d \overline{{\cal B}_{lm}} \overline{C}_m \,+\, \sum_m \overline{D}_{lm} 
\overline{C}_m dt) 
\nonumber \\
\ &\, \nonumber \\
\ &\,\ +\, \sum_{m m'}  d \overline{{\cal B}_{lm}} d {\cal B}_{lm'} \,+\, O(dt^2)   
\end{array}
\label{eq:martin0}
\eeq
\noindent where we used Eq. (\ref{eq:finalB}). To get the variation of the norm one has to sum over $ l $. The terms linear in $ d{\cal B} $ cancel each others and separating the diagonal terms $ l = m $, one gets
\begin{eqnarray}
\sum_l d(\overline{C}_l C_l) &\,=\,& \sum_{l\neq m} \big(\, \overline{C}_l D_{lm} C_m + C_l \overline{D}_{lm} \overline{C}_m + F_{lm}^2 \vert C_m \vert^2 \, \big) dt \nonumber\\
\ &\, & \nonumber\\
 \,  &\,+\,& \sum_m \big(  D_{mm}  +  \overline{D}_{mm} + F_{mm}^2\big) \vert C_m \vert^2 dt \,=\, 0 
\label{eq:consnorm}\end{eqnarray}  
\noindent This equation must be satisfied identically for any possible evolution of the system, i.e. all possible set of coefficients $ C_l $, compatible with a fixed norm, and for a generic system. The first requirement implies that the matrix 
$ D_{lm} $ must be a fractional polynomial of the coefficients of order zero, since a polynomial can vanish identically only if each monomial term is set to zero. The second requirement implies that if for a couple $(l,m)$ one has $ F_{lm} = 0 $ also the corresponding matrix element $ D_{lm} = 0 $, since the stochastic coupling between the states $l$ and $m$ vanishes. Then the simplest solution for the matrix $ D $ is
\beq
D_{lm} \,=\, - \frac{1}{2 \overline{C}_l} F^2_{lm} \overline{C}_m
\label{eq:matD}\eeq 
\noindent In principle higher powers of the $ C $ are possible in the numerator and in the denominator, but we assume that the simplest solution is the correct one (Occam's razor). The stochastic equation (\ref{eq:rightone}) is then determined
\beq
\imath d C_l(t) \,=\, \sum_m d {\cal B}_{lm} C_m(t) - \frac{\imath}{2\overline{C}_l}\sum_m F_{lm}^2 \vert C_m \vert^2 dt
\label{eq:deter}\eeq
\noindent Notice that it is nonlinear. It implements the Schrodinger equation, embodied in the interaction representation, adding a stochastic component in the evolution of the system wave function. It is a direct consequence of the conservation of the norm of the standard component, as demanded by the assumption that the wave function is representative of the state of a given system, in agreement with ordinary QM.\par 
If one introduces explicitly the modulus and phase of the component $C_l$, this equation coincides with the stochastic equation of ref. \cite{Pearl1,Pearl2}, where it was postulated without specific justifications.\par 
A fundamental consequence of the explicit expression of the matrix $ D $ is the stochastic equation for the modulus square of each single component $ C_l $, which reduces to
\beq
d(\overline{C}_l C_l) \,=\,  - \imath \sum_m d {\cal B}_{lm} C_m \,+\, \imath \sum_m d \overline{{\cal B}_{lm}} \overline{C}_m 
\label{eq:single}\eeq
\noindent Because of Eq. (\ref{eq:av0}), if we now consider the average value over the stochastic trajectories, one gets
\beq
< d(\overline{C}_l C_l) >\,=\, 0
\label{eq:martin}\eeq
\noindent which is usually referred as 'martingale' property. In addition, in agreement with ref. \cite{Pearl1,Pearl2}, one can
calculate the stochastic differential of the product of the modulus square of two components. One finds
\beq
d < (\vert C_l\vert^2 \vert C_m \vert^2) > \,=\, - F_{lm}^2 < \vert C_l\vert^2 \vert C_m \vert^2 > dt
\label{eq:cl2cm2}\eeq
\noindent which implies a monotonic decrease to zero of each $ \vert C_l \vert^2 $, except at most one. This is reduction.
\section{Answer to the Questions. \label{sec:aquestions}}
In this section we discus how the model is able to give an answer to each one of the question posed in Section \ref{sec:questions}. 
\vskip 0.3 cm
\par\noindent
{\it Reduction.} 
\vskip 0.3 cm
\par 
In summary, the results of the last Section imply the following behavior of the stochastic evolution.
If the initial state is a superposition of a set of states 
\beq
 \vert \Psi (0) > \,=\, \sum_l \vert \Psi_l > C_l(0)
 \label{eq:insup}
\eeq
then, during the stochastic evolution
\vskip 0.2 cm
\par\noindent
1. The summation of the square amplitudes $ \sum_l  \vert C_l(t) \vert^2 $ remains constant for each trajectory, 
so if it is initially equal to 1, it remains 1. This is a basic physical requirement.
\vskip 0.3 cm
\par\noindent
2. The final amplitudes tend to zero except only one (with a modulus square equal to 1), i.e. a single state 
$ \vert \Psi_l > $ is dominant after a interval of time longer than $1/\vert F_{lm} \vert^2 $. 
\vskip 0.3 cm
\par\noindent
3. Each state $ \vert \Psi_l > $ becomes dominant with a relative frequency equal to the initial modulus square $ \vert C_l(0)  \vert^2 $ (Born' s rule).
\vskip 0.2 cm
\par 
\noindent
This behavior entails a reduction of the wave function, i.e. it reproduces the fundamental properties of what is usually assumed to occur in Q.M. measurement process, whatever meaning is given to this word.
\par 
The obvious question is when and under which conditions the reduction takes place. First of all the interaction strength between the nonstandard parts of two states must be infinitesimal, since it involves unlimited wave numbers and frequencies. If the remaining part of the matrix elements of the interaction is finite, the coupling between the nonstandard parts has therefore no effect on the temporal evolution. The relevant case is when one considers the matrix elements between the nonstandard components of the same state, i.e. in a quantum superposition, provided they become unlimited enough to overcome the infinitesimal character of the interaction strength. \par 
Let us consider a schematic representation of a generic measurement. If initially the system is in a superposition of eigenvalues $ \{a\} $ of the observable $ A $ to be measured, initially each component will evolve at the standard level    
according to the response of the detector to that particular eigenvalue, different for each one, and their superposition remain intact (linear evolution). If the detector is a proper one, the evolution must bring the responses towards a macroscopic level, i.e. the number of excited degrees of freedom must increases as the standard dynamics evolves. 
In Appendix C it is shown that the matrix elements between the components of the superposition scale with the number of excited degrees of freedom $ N_d $ as $ \lambda^{2(N_d - 2)} $, where $ \lambda $ is the unlimited number of nonstandard wave vectors. The factor $ F_{lm} $ of Eq. (\ref{eq:finalB}) is given by
\beq
F_{lm} \,=\, \sqrt{a_0 N_\Omega} v_0 \lambda^{2(N_d - 2)} I_{lm}
\label{eq:Flm}
\eeq
\noindent where $ v_0 $ is the infinitesimal strength of the nonstandard interaction. The value of $ N_d $ is increasing with time towards macroscopic values. This estimate is obtained in the case of position measurement, but it is similar in other cases. \par 
The reduction takes place at some critical value of $ N_d $, when the factor (\ref{eq:Flm}) passes from an infinitesimal value to an unlimited value. This occurs in an interval of time necessary to increase by one unit $ N_d $,
which is of the order of the standard interaction time between the system and a single degree of freedom, e.g. the time necessary to excite a molecule. In fact the interaction cannot be strictly instantaneous, and it is only within this uncertainty that the time for the reduction process can be defined. Notice that the reduction is complete, i.e. the components other than the reduced one have infinitesimal probability, which is zero at standard level. 
Although this result could appear not physical, it has been a standard assumption in ordinary quantum mechanics
that the "quantum jumps" are indeed instantaneous, which is also compatible with the observational indications \cite{jump1,jump2,jump3}. The model gives a direct explanation of these phenomenological findings. \par 
The model, as developed here, cannot predict the critical value of $ N_d $, but the exponential growth of $ F_{lm} $
ensures that it exists. In fact, according to Eq. (\ref{eq:upper}), the factor with the square root can be diverging at most as $ d^{-\beta/2} $, with $ \beta < 1 $. On the other hand, although the extrapolation of the effective interaction $ v_0 $ to the nonstandard sector can be uncertain, it is legitimate to assume that the interaction mechanism is the same as in the standard sector, i.e. it is mediated by the exchange of the quanta of the corresponding field, as photons, gluons, mesons. In this case  
the effective strength is damped by the quanta propagator, which is suppressed at least by a factor $ 1/k^2 $, with $ k $
the unlimited wave vector transfer. 
Then, according to Eq. (\ref{eq:range}), the strength $ v_0 $ is infinitesimal at least as $ d^{2\delta} $. In order that this mechanism is actually working one has to impose the constraint $ \delta > \beta/4 $. 
In addition, if one assumes $ \lambda \sim d^{-\gamma} $, only for an exceedingly small value of $ \gamma $ the critical $ N_d $ can exceed large macroscopic values. These are the boundaries that the structure of the model are imposing to the parameters that cannot be specified precisely. 
\par 
The stochastic equations (\ref{eq:rightone}),(\ref{eq:matD}) are non-linear and therefore it is necessary to select the representation in which they are assumed to hold. In a measurement it is the detector that performs the selection. In fact 
the response of the apparatus occurs only for a definite set of initial states of the system to be detected, and therefore the evolution of the system + detector can be only along a linear combination of the responses, whatever is the initial state of the system. To exemplify, let us consider the measurement of the position of a particle by a scintillator in a schematic picture. The only possible response of the detector to the impinging particle is the excitation of a certain number of atoms or molecules, which are each one well localized. Then a particle described by a localized wave packet will excite the molecules along its path in the detector. A superposition of wave packets will produce the corresponding superposition of such  responses. 
For any initial state, the response of the detector must take place within the space spanned by these possible localized responses, which selects uniquely the representation that must be considered. No other response of the detector is possible. This also specifies the meaning of an "ideal detector". 
\par 
In the derivation of the reduction process infinitesimal and unlimited numbers have been used. This can appear puzzling,
but it has to be stressed that these numbers are legitimate as the others, once \Rns\, has been introduced. 
\par
Finally let us consider the case in which the detector response is too weak to reach the macroscopic threshold. Then the detector will keep the coherent superposition of the different responses. However, to observe the result of the measurement one could use a probe, which can act as a more sensitive detector and the reduction will occur. As a schematic typical case one can consider a photographic plate, where the silver grain will appear once the photographic process is performed. 
As a limiting case the probe can be eventually our physiological perception (not to be confused with consciousness !), which necessarily involves macroscopic systems.
\vskip 0.3 cm
\par\noindent
{\it General reduction and the classical limit.}
\vskip 0.3 cm
\par 
Within the model the reduction process cannot be restricted to the case of measurements, but it should occur in a generic physical systems. It is therefore appropriate to clarify its role in the structure and dynamics of a system.
The reduction must take place within the states that are available under the physical conditions of the system,
which fixes e.g. the conserved quantum numbers. If the ground state is not degenerate it will be the only available
state. As the average energy of the system increases, more an more excited states will be available. If the system has a number of degrees of freedom below the 'macroscopic' threshold the dynamics will develop without any reduction process and the standard quantum treatment will be the appropriate one. If the threshold is overcome a modification of the dynamics will occur. In this case no coherent superposition can be present. In a given representation 
$ \{a_j\} $ the state of the system at the initial reference time $ t = 0 $ must be described by a diagonal density matrix 
\beq
\rho(0) \,=\, \sum_j | a_j > p_j < a_j |
\label{eq:denmat}\eeq
\noindent where $ p_j $ is the probability that the system is in the pure quantum state $ \{a_j\} $, with 
\beq
\sum_j p_j \,=\, 1
\label{eq:probtot}\eeq 
\noindent During the evolution the density matrix remains diagonal
\beq
\rho(t) \,=\, \sum_j U(t)| a_j > p_j < a_j |U(t)^{\dag} \,=\, | a_j(t) > p_j < a_j(t) |
\label{eq:denmat1}\eeq
\noindent However we need the density matrix in the basic states of the representation
$ \{a_j\} $, under the condition that no coherent superposition is present among them, i.e. the density matrix is still diagonal. Non-diagonal matrix elements imply the appearance of coherent superposition between basis states of the representation, which will be instantaneously canceled by reduction 
\beq
 \rho(t)  \,=\, \sum_j | a_j > p_j(t) < a_j | 
\label{eq:orig}\eeq
\noindent with
\beq
p_j(t) \,=\, \sum_k | < a_j | a_k(t) > |^2 p_k
\label{eq:ptime}\eeq 
\noindent It is readily verified the conservation of the trace
\beq
Tr(\rho(t)) \,=\, \sum_j p_j(t) \,=\, \sum_j p_j \,=\, 1
\label{eq:tracet}\eeq
\par 
Now the stochastic equations (\ref{eq:rightone}),(\ref{eq:matD}) must have the same form in any representation, since they are dictated by the conservation of the norm. This formal invariance indicates that there is no simple analytical relation between the non-linear terms of the equations in different representation.
As a consequence of the invariance the expression (\ref{eq:orig}) must be valid in a generic representation. This conclusion can be valid under the hypothesis of \textit{random a priori phases} \cite{Pathria}, so that changing representation the summation over the phases averages to zero the off-diagonal terms and the density matrix remains diagonal. As a consequence the different representations of the evolution equation (\ref{eq:orig}) can be considered equivalent since they can be related effectively by a unitary transformation. The random a priori phases hypothesis is actually essential in statistical thermodynamics.
Indeed for an isolated system in thermodynamic equilibrium in the microcanonical ensemble the statistical density matrix is diagonal in the energy representation \cite{Pathria} and it can be written (not normalized)
\beq
\rho(E) \,=\, \sum_i | E_i > < E_i |
\label{eq:microc}\eeq
\noindent where the energies $ E_i $ are restricted to a narrow interval around the reference energy $ E $. The diagonal form is valid in other representations
provided the random a priori phases hypothesis is adopted \cite{Pathria}. In this case the system is ergodic, i.e. all the states of the representation has the same probability to be occupied during the time of the observation of the system, as it is assumed in Eq. (\ref{eq:microc}), but the suppression of the off-diagonal elements holds in general. Notice that if the initial $p_k$ are all equal, they remain equal for all time. In the thermodynamic limit the hypothesis of random phases looks more likely, but in any case even for a moderately macroscopic system the number of microstates can be exceedingly large, which supports the assumption.
If this assumption is not strictly valid, the different representation are not equivalent, and the dynamics will depend on the representation in an essential way. In this case it would be necessary to describe the system by a multi-valued state vector. 
\par 
If in all representations the density matrix can be taken diagonal, the description of the system looks similar to a classical description, where one can fix the values of all dynamical variables of a system at the same time, irrespective
to their possible quantum incompatibility. Of course this is not enough, one has to demand that the evolution of each variable to be close to the classical one. This is possible only for the physical quantities for which the corresponding
action is exceedingly large with respect to $ \hbar $, i.e. in the limit of short standard wavelengths. In this case the superposition among different values of a given variable is suppressed by reduction and the classical value will be overwhelming dominant. As an example one can consider the center of mass of a macroscopic object, for which one can construct a narrow wave packet because of the short wavelengths associated with a large mass. If the system is macroscopic, the diagonal form of the density matrix and the corresponding reduction processes will hinder the spread of the wave packet beyond the Erhenfest time, so that the quantum description can merge in the classical one. 
\par 
Another relevant case is when two distinct systems weakly interact in the standard sector. One can describe the overall system by the superposition of product states, within the states available according to the physical conditions, e.g.the total energy. For macroscopic systems the reduction process will select one of the product state at each instant of time.
If the distance between the two systems increases, the available states will reduce, until only one product state is 
possible. This is the mechanism of decoupling between tow systems that are moving apart.
\par 
The Schroedinger cat paradox finds a simple solution, because no superposition between "dead" and "alive" states of the cat can appear, since it will be canceled by reduction. Actually the reduction takes place already during the event of decay of the radioactive nucleus, which is a stochastic process. A more detailed description of the evolution of the "experiment" requires a closer analysis of the decay mechanism and dynamics, which is outside the scope of the present paper.
\vskip 0.3 cm
\par\noindent
{\it Spontaneous decay.}
\vskip 0.3 cm
\par 
In Section \ref{sec:questions} we have pointed out the question raised by the spontaneous emission, in particular of excited atoms. Before considering the problem within the model, we need first to recapitulate how the decay process can be treated in standard QM. Let us consider a two level atom excited to the upper level that evolves in time according to the ordinary QM.
The hamiltonian can be written as the sum of the atomic part, the photon part and the electromagnetic coupling
\beq
H \,=\, H_a \,+\, H_{ph} \,+\, v_e
\label{eq:ham}\eeq
\noindent
and the basis states are the atomic excited state $ | E_2 > $ and the atomic ground state plus a photon 
$ | E_1 E_{\gamma} > $,
with their corresponding energies and projection operators
\beq
P \,=\, | E_2 >< E_2 |  \ \ \ \ \ , \ \ \ \ \ Q \,=\, \int_{E_\gamma} | E_1 E_{\gamma} >< E_1 E_{\gamma} |
\label{eq:proj}
\eeq
The decay amplitudes of the excited state $ | E_2 > $ are the probability amplitudes $ A_2(t) $ to remain in such excited state at the time $ t $ and the complementary probability amplitudes $ A_\gamma(t) $ to decay into one of the states $ | E_1 E_{\gamma} > $
\beq
A_2(t) \,=\, < E_2 | \exp(-\imath H t) | E_2 > \ \ \ \ \ \ , \ \ \ \ \ \ A_\gamma(t) \,=\, < E_1 E_{\gamma} | 
\exp(-\imath H t) | E_2 >
\label{eq:probA}\eeq
\noindent
These matrix elements of the evolution operator can be obtained by contour integration in the complex energy plane of the 
corresponding matrix elements of the resolvent operator
\beq
A_2(t) \,=\, -\frac{\imath}{2\pi}\oint d\omega  < E_2 | \frac{1}{\omega - H + \imath\eta} | E_2 > 
\label{eq:contour}\eeq
\noindent
where the integration is in the lower complex plain, along the usual semi-circle, and $\eta$ an arbitrarily small quantity. Analogous expression holds for $ A_\gamma $.
Using the properties of projection operators, one can derive the well known identities
\begin{eqnarray}
P \frac{1}{\omega - H} P &=& \Big[ \omega - P H P - P v_e Q \frac{1}{\omega - Q H Q} Q v_e P \Big]^{-1} 
\nonumber \\
\ &\, \nonumber \\
Q \frac{1}{\omega - H} P &=& \frac{Q}{\omega - Q H Q} v_e P \frac{1}{\omega - H} P 
\label{eq:iden}\end{eqnarray}
\noindent
It follows that the resolvent amplitude $ A_2(\omega)$ pertinent to $ A_2 $, appearing in Eq. (\ref{eq:contour}), can be written
\beq
A_2(\omega) \,=\, \Big[ \omega - E_2 + \int_{E_\gamma}\frac{ | v |^2 }{\omega - E_1 - E_\gamma + \imath\eta} + \imath\eta \Big]^{-1}
\label{Aomega}\eeq
\noindent
where $ v \,=\, < E_1 E_{\gamma} | v_e | E_2 > $. The integral can be split into real and imaginary part by using well known properties of the Dirac delta-function. One gets
\beq
A_2(\omega) \,=\, \Big[ \omega - E_2 + \mathcal{P}\int_{E_\gamma}\frac{ | v |^2 }{\omega - E_1 - E_\gamma} + 
\imath\pi | v_1 |^2 \rho \Big]^{-1}
\label{eq:split}\eeq
\noindent where $ \mathcal{P} $ denotes principal part of the integral, $ v_1 $ is the interaction matrix element calculated at $ \omega - E_1 $ and $ \rho $ the corresponding density of states. This expression indicates the presence of a pole in the lower complex energy plane. For the electromagnetic transitions the weak coupling regime is usually valid. Then in first approximation the real part of the position of the pole can be taken as the value of $\omega$ where the real part of the denominator vanishes and the imaginary part can be calculated at that value. Then the contour integral of Eq. \ref{eq:contour}) can be performed using the residue theorem. One gets
\beq
| A_2(t) |^2 \,=\, \exp( -\Gamma t )
\label{eq:A2}\eeq
\noindent where $ \Gamma \,=\, 2\pi| v_1 |^2 \rho $ can be interpreted as the probability for unit of time for the atom to decay to the ground state. Notice that, according to standard QM, the expression of Eq. (\ref{eq:A2}) can be interpreted as the probability for the atom to remain in the excited state, but the wave function continues actually to evolve until a measurement is done and the transition can occur. 
\par 
A similar procedure can be followed for $ A_\gamma(t) $, using the second of the identities (\ref{eq:iden}). One gets
\beq
| A_\gamma(t) |^2 \,=\, | v_1 |^2 \frac{1 + \exp(-\Gamma t) - 2\exp(-\Gamma t/2) \cos(\Delta E t)}{\Delta^2 + \Gamma^2/4}
\label{eq:Agamma}\eeq
\noindent where $ \Delta E \,=\, \omega_0 -E_1 - E_\gamma $, with $ \omega_0 \approx E_2 $ the real part of the pole. The probability to decay into one of the state $ | E_1 E_\gamma > $ has the expected Lawrencean shape
at any time, with some time dependent modulation. One can verify explicitly that 
\beq
\int_{E_\gamma} | A_\gamma(t) |^2 \,=\, 1 \,-\, \exp(-\Gamma t)
\label{eq:cons}\eeq
\noindent as demanded by conservation of probability.\par
It has to be stressed that this treatment includes the coupling $ v_e $ to the electromagnetic field to all orders.
The result is in agreement with ref. \cite{PRA}, where second order perturbation theory was used. The latter was interpreted as describing the "radiation reaction" triggering the spontaneous emission, or, alternatively, as the effect of vacuum fluctuations. Actually, both interpretations correspond to radiative corrections of Quantum Electrodynamics. The fact that we approximate the position of the pole at the unperturbed energy is equivalent to neglect the Lamb' s shift.\par 
Despite these quantum mechanical results appear consistent with expectations, as explained in detail in Section \ref{sec:questions}, they are difficult to reconcile with phenomenology. This difficulty is overcome in ordinary QM by an additional assumption. As written in any QM textbook, one assumes that the atom can decay randomly with constant probability per unit of time.  This probability can be calculated in the long time limit and it coincides with $ 1/\Gamma $, consistent with the above QM calculation. This assumption leads immediately to the expected exponential law for the detection probability, independently of the distance of the detector, consistent with phenomenology.
Equivalently, a set of decaying atoms, excited at the same initial time, will show a population of excited atoms that decays exponentially with time, again consistently with elementary observations. This assumption is not part of the standard QM basic formulation, and it introduces a stochastic evolution, the so called "quantum jumps", not included in the standard formalism. Of course quantum decay has been present since the dawn of QM, but the origin of their stochastic character has not yet clearly tracked down. In general one invokes the vacuum fluctuation as a trigger of the quantum jumps corresponding to the decay processes, but an explicit description of the jump is absent. We will now show that the nonstandard component of the vacuum fluctuations can indeed produce the stochastic decay process.
\par 
As it is well known the state with no photon, that is the quantum state where all modes are in their lowest level of $ \frac{1}{2} \hbar \omega $, is not the true quantum vacuum. In fact in the vacuum state there must be a continuous creation and annihilation of electron-positron pairs. As a consequence the electromagnetic field must fluctuate in time. 
The standard part of the vacuum field is detectable indirectly, since it is responsible of e.g. the Lamb' s shift and of the Casimir effect. In the elementary treatment of the Lamb' s shift \cite{Itzykson-Zuber} the interaction with the vacuum field introduces oscillations of the electrons of all standard frequencies, within a cut-off interval, which in turn results in an energy shift on the bound electron. The typical time scale of the fluctuations can be estimated from the uncertainty principle. If a fluctuation of energy $\Delta E $ occurs, it will last for a time $\tau \sim \hbar/\Delta E $. Taking for 
$\Delta E $ the energy necessary to create an electron-positron pair, i.e. about 1 MeV, one gets a $\tau$ of the order of $10^{-21}$ sec., which is much shorter of a typical electromagnetic process. These fluctuations are expected to occur with a frequency uniform in time. 
One has to
notice that the decay process is equivalent to a reduction of the quantum state. In fact at the initial stage of the evolution the quantum state is just the linear superposition of the excited state
$ | E_2 > $ and the ground state plus the emitted photon $ | E_1 E_{\gamma} > $ and the decay is just the reduction of the superposition to the second component. The reduction cannot be triggered by a standard fluctuating field, of whatever origin, since the standard evolution keeps the superposition among the possible outputs and no selection, i.e. quantum jump, can occur. Indeed it can give a contribution only to the Lamb's shift.
\par 
The field fluctuation has a standard part $\phi^S(t,x_j)$ and a corresponding nonstandard part $\phi^{NS}(t,x_j)$, given by
\beq
\phi^{NS}(t,x_j) \,=\, \sum_{k_v} \exp(\imath k_v x_j) \cos(\omega(k_v) t) \phi^S(t,x_j)
\label{eq:fluNS}\eeq
\noindent in line with the nonstandard part of a generic wave function, and indeed it satisfies the nonstandard wave equation within the same criterion. For a fluctuation around the atom, this nonstandard 
field will act on the electron and will couple the non decayed and the decayed states with the matrix element
\begin{eqnarray}
&A_{21}&\!\!\! =\, < E_2 | \phi^{NS}_v | E_1 > \,=\,  C \sum_{x_j \{k\}}\ d\ \overline{\psi_2^{S}}(t,x_j) \psi_1^{S}(t,x_j) \phi^S(t,x_j) \nonumber \\
 &\times&\!\!\!\!\! \cos(\omega(k_1)t) \cos(\omega(k_2)t)\cos(\omega(k_v) t) \exp(\imath(k_1 - k_2 + k_v) x_j)\ 
\label{eq:nsme}\end{eqnarray}
\noindent where $\psi_1^{S}$ and $\psi_2^{S}$ are the standard parts of the wave functions for the states $| E_1 >$ and
$| E_2 >$ and $ C $ is a finite factor which includes the electron charge. The factor in the first line 
$Z(t,x_j) = \psi_1^S \psi_2^S \psi_1^{S}$ is the canonical extension of a standard function, and therefore its expansion
\beq
Z(t,x_j) \,=\, \sum_q \tilde{Z}(q,t) \exp(\imath q x_j)  
\label{eq:sexpa}\eeq
\noindent contains only finite wave vectors $q$. Insertion of (\ref{eq:sexpa}) in (\ref{eq:nsme})and the summation over $x_j$ gives
\begin{eqnarray}
A_{21} \,=\, C \sum_q \sum_{\{k\}}  N \tilde{Z}(q) \cos(\omega(k_1)t) \cos(\omega(k_2)t)
\nonumber \\
\times \big(\exp(\imath\omega(k)t) \,+\, C.C. \big)
\label{eq:intxj}\end{eqnarray}
\noindent where $ k = k_2-k_1-q $ and $C.C.$ means complex conjugate.
 The summation over $ q $ can be performed after the expansion in $ q $ of ($\Delta k = k_2 - k_1$)
\beq 
\omega(\Delta k - q) \,=\, \frac{2}{d} \sin((\Delta k - q)/2N)  
\label{eq:omegaq}\eeq
\noindent
as follows
\beq
\omega(\Delta k - q) \,=\, \frac{2}{d}[\,\sin(\Delta k/2N)\cos(q/2N) \,+\, \cos(\Delta k/2N)\sin(q/2N)\, ] 
\label{eq:qexp}\eeq
\noindent Neglecting infinitesimal corrections one gets
\beq
\omega(\Delta k - q) \,=\, \omega(\Delta k) \,+\,   q v
\label{eq:finq}\eeq
\noindent where $ v = \cos(\Delta k/2N) $ is the group velocity of the corresponding mode. Then the summation over $ q $ gives 
\beq
A_{21} \,=\, C N \sum_{\{k\}} H(t) \cos(\omega(k_1)t) \cos(\omega(k_2)t)
\label{eq:fin}\eeq
\noindent
where 
\beq
H(t) \,=\, Z(vt)\exp(\imath\omega(k_2-k_1)t) \,+\, Z(-vt)\exp(-\imath\omega(k_2-k_1)t)
\label{eq:Hfac}
\eeq
\noindent where the $Z$ functions vary over a standard time scale, and therefore they can be taken constant in each monad.
 As expected the result is a summation of an unlimited number of nonstandard frequencies. As in the general case, Eq. (\ref{eq:fin}) entails a stochastic process. In this case the factor $F_{12}$ in front of the stochastic process 
 $ {\cal B}(t) $ turns out to be $N\sqrt{N_\Omega a_0}$, which is unlimited. The evolution equation is therefore singular, or ill defined, around the time $t_0$ of the beginning of the fluctuation. A regularization is necessary, which must project the dynamics from \Rns\ to \Rs\ . The regularization can be achieved simply by replacing the constant factor $F_{12}$ by a factor $G_{12}$ which grows from a finite value to an arbitrary large value in a finite time interval $\Delta t_g$. For each value of $\Delta t_g$ the reduction will take place within this time interval. At decreasing value of $\Delta t_g$ the reduction of the wave function will occur in a shorter and shorter interval of time. The final choice between the excited and decayed state of the atom can be different, but the Born's rule will be always satisfied. In this sense one can speak of a stable limit of the reduction as it occurs in an arbitrarily small time in \Rs\ .
This result is all what we need for the description of the stochastic decay. \par 
If $ | E_2 > $ is the initial state, in a generic case it will evolve according to the standard dynamics, and at the time  $ \Delta t_1 $ the amplitude square for the $ E_2 $ component will be $ \exp(-\Gamma \Delta t_1 )$. If at that time a vacuum fluctuation occurs, the system will remain in $ E_2 $ indeed with probability  $P_1 = \exp(-\Gamma \Delta t_1) $, or it will decay with the probability $ 1 - P_1 $. In the latter case the instantaneous  decay has occurred and the evolution of the electron stops. In the former case the process will be repeated within a time interval $ \Delta t_2 $ and the overall probability to remain still in $ E_2 $, i.e. not decayed, will be
\beq
P_2 \,=\, \exp(-\Gamma \Delta t_1) \exp(-\Gamma \Delta t_2) \,=\, \exp(-\Gamma t_2)
\label{eq:2step}\eeq
\noindent with $ t_2 = \Delta t_1 + \Delta t_2 $. The evolution will continue, alternating standard and stochastic dynamics, until the decay occurs. After $ n $ steps, the probability to have not decayed will be
\beq
P_n \,=\, \exp(-\Gamma t_n)
\label{eq:nstep}\eeq
\noindent with $ t_n = \sum_i \Delta t_i $. Since the time intervals $ \Delta t_i $ are much shorter than $ 1/\Gamma$,
the evolution will be practically continuous. In this way one recover the exponential law for the probability to have not decayed and the constancy of the probability per unit of time to decay, as demanded by phenomenology. Of course we have implicitly assumed that the time scale for the occurrences of vacuum fluctuations is not too large, otherwise the exponential law would be affected. This is in line with the interpretation assumed in ordinary QM.\par 
In summary, while the standard part of the vacuum fluctuations are responsible for the Lamb's shift and of the Casimir effect, the corresponding nonstandard component is responsible for the ''spontaneous" electromagnetic decay. Both components cannot be directly detected.
\vskip 0.3 cm
\par\noindent
{\it Reduction with no interaction.}
\vskip 0.3 cm
\par 
If a set of identical atoms in their ground state are excited by a photon beam, one observes that the number of excited atoms increases exponentially in time at a well defined rate. This observation indicates that the absorption of a photon by a single atom is a random process with a constant probability per unit of time. Also in this case one has to introduce a stochastic process which is beyond the standard QM formalism. The excitation of a single atom can be described along the same lines as for the decay. Before the quantum jumps, the excitation dynamics is just the time reversal of the decay evolution. The vacuum field fluctuations at a certain stage produce the stochastic excitation process, which entails a probability that increases exponentially. \par     
In the schematic example in Sec. \ref{sec:questions} of a null experiment with the Mach-Zender interferometer, let us suppose for simplicity that the obstacle in one of the branch is just a single atom. The relevant wave function $\Psi_i$ of the system at the exit from the first splitter can be written 
\beq
| \Psi_i > \,=\, \frac{1}{\sqrt{2}} \big( | \gamma_1 > \,+\, | \gamma_2 > \big) | E_1 >
\label{eq:split1}\eeq
\noindent where, as before, $E_1$ indicates the ground state of the atom, and $\gamma_1$ and $\gamma_2$ the components of the photon wave packet in the branches with and without the atom, respectively. When the photon arrives at the atom, before the quantum jump occurs, the wave function will be
\beq
| \Psi > \,=\, \frac{1}{\sqrt{2}} | \gamma_2 E_1 > \,+\, \frac{1}{\sqrt{2}} \big( C_1 | \gamma_1 E_1 > + C_2 | E_2 > \big)
\label{eq:befj}\eeq 
\noindent where $E_2$ is the excited state, and 
\beq
| C_1 |^2 \,=\, e^{-\Gamma t} \ \ \ \ ; \ \ \ \  | C_2 |^2 \,=\, 1 - e^{-\Gamma t}
\label{eq:Ccoeff}\eeq
\noindent with $\Gamma$ the transition rate. If the latter is large enough, in short time the amplitude $C_1$ will be negligible and the wave function can be written
\beq
\frac{1}{\sqrt{2}} \big( | \gamma_2 E_1 > \,+\, | E_2 > \big)
\label{eq:late}\eeq
\noindent The quantum jump can occur at this stage, due to the stochastic coupling between the two components, in the same way as in the decay process. Then there will be 50\% probability that the photon is absorbed and 50\% probability that the atom remains in the ground state and the photon $\gamma_2$ passes undisturbed. Taking into account that the port has been arranged in the dark configuration, the overall probability for the photon to be detected there turns out to be
$ 1/4 $. In the latter case one has the 'paradox' that the reduction appears to occur without interaction. This is the physical situation that occurs when there is a completely absorbing object in place of the atom. In fact in this case the rate of absorption $ \Gamma $ is so large that amplitude of the ground state component goes to zero in a vanishing small time.\par 
In the general case, e.g. with a semi-transparent object, one has to consider the reduction of the full wave function (\ref{eq:befj}). Since there is no stochastic coupling between the states $| \gamma_1 E_1 >$ and $| \gamma_2 E_1 >$, the only coupling is between the states $| E_2 >$ and $\frac{1}{\sqrt{2}} ( | \gamma_2 E_1 > \,+\,  C_1 | \gamma_1 E_1 > )$. Then the reduction will give the excited state with probability $ |C_2 |^2/2 < 1/2 $ or the other state with probability $
  1 - | C_2 |^2/2 $, where $ C_2 $ is taken at the moment of the quantum jump. At the final port of the interferometer, that is arranged in the dark configuration, the probability to detect the photon turns out to be
$ |1 - C_1 |^2/4 $. If finally $ C_1 = 1 $ means that the photon has not been absorbed by the object, while for $ C_1 = 0 $ we get the result of the previous case. The probability to detect the photon in the orthogonal state , e.g. at the other port, is in general $ |1 + C_1 |^2/4 $. Summing up the probabilities of the three possible outputs (detection at the originally dark port, detection at the other port and absorption by the obstacle) one gets in fact $ 1 $.  
\section{Summary of the model and conclusions.\label{sec:summodel}}
We first summarize the steps that lead to the formulation of the model and the basic phenomena that it can describe.
\par\noindent 
1. The time axis and each space coordinate axis are extended from the standard real axis \Rs\ to the nonstandard one \Rns. This step does not require any particular assumption, because nonstandard analysis is a well established mathematical theory. Accordingly, the wave functions acquire a nonstandard component.
\par\noindent
2. For each space coordinate the hyperfinite real line is introduced, with infinitesimal lattice spacing. The same is done for the time axis, but with an infinitesimal lattice spacing of higher order. This fixes the manifold where physical phenomena occur and the support of the wave functions. The same type of time and space lattices have been already introduced in ref. \cite{Kass}.
\par\noindent
3. The wave equation is solved in the extended space-time manifolds. The nonstandard part
of the wave function turns out to be necessarily time dependent. As a consequence, the nonstandard frequencies and wave vectors cannot be related to energy and momenta, respectively.
\par\noindent
4. The interaction matrix elements between the nonstandard parts of the wave function components for a given physical system are characterized by unlimited frequencies. Since the nonstandard time scale is not observable, one must resort to a coarse graining and averaging procedures at the level of each monad of time. One can demonstrate that this automatically implies a white noise and an associated Ito stochastic process for the matrix elements in \Rs.
\par\noindent
5. By imposing the conservation of the state norm, i.e. the norm of the standard part, one gets the explicit stochastic equation for the wave function components, which implements the standard Schr\"odinger equation. With a suitable choice of the normalization of the nonstandard component, this equation entails the reduction process and the Born's rule without further assumptions.
\par\noindent
6. A similar procedure explains the stochastic character of the spontaneous decay of a system by photon emission and the 'paradox' of the null experiments.
\par 
Points 4 and 5 are the basic steps in building the model. In deriving the reduction process at the step 4, one finds that there must be a critical value of the "size" of a system, i.e. of the number of independent degrees of freedom, above which no coherent superposition is possible. This fixes the transition from quantum to classical physics. Unfortunately the model is not able to fix the critical size, which has to be fixed by phenomenology. The description of the dynamics in this regime is unique
if one adopts the "random a priori phases" assumption, which is mandatory in quantum statistical theory.
\par
In conclusion, in this paper I have proposed a model that extends the formalism of QM in order to include the observed physical phenomena, like the measurement processes and the quantum jumps in an atomic decay process, that are not described by the standard QM formalism and which appear just as postulates in the standard theory. The stochastic, irreversible and indeterminacy characters of the quantum processes are the consequences of a basic stochastic dynamics which arises naturally once the proper nonstandard manifold is introduced. The model is intended to be the basic scheme for a possible more refined theory. 
\section{Final considerations.}
The question that naturally arises is twofold : i) what the model is able to achieve ? ii) is there any possibility to check the model and eventually are there predictions of the model that can be verified ?\par
First of all the model was aimed to explain some basic phenomenology of Quantum Mechanics that lacks explicit formal description, as discussed in the Introduction. To the knowledge of the author no other model is able to answer, within the same scheme, all the questions formulated in the Introduction. This is the main scope of the model. \par 
To illustrate and stress this point, it is useful the comparison with other models that are aimed to implement quantum mechanics formalism, in particular for the measurement problem. One of the most known and discussed model is the "Continuous 
Spontaneous Localization" (CSL) model \cite{CSL_RMP}, which postulates the presence of an additional process, of unknown origin, where the wave function undergoes an instantaneous space localization, following the probability Born' s rule. The rate of the localization is such that it can occur mainly for macroscopic system. This can explain the wave function reduction in position measurements and the classical limit for the center of mass motion of macroscopic objects. The model modifies the evolution of the wave function as described by e.g. the Schrodinger equation, by an additional term which can be put in the form of Lindblad
master equation \cite{Lindblad}. The localization process should produce some energy dissipation. Several experiments have tried to detect the effect of the Lindblad \cite{qtest} term and the dissipation, with no compelling evidences.  
It has to be noticed that CSL cannot explain the null experiments, as well as spontaneous emission. Furthermore CSL deals only with reduction of the wave function in position and it cannot predict reduction in other physical observables.\par 
It has to be stressed that the model developed here predicts that the wave function reduction can occur only if the number of independent degrees of freedom exceed some macroscopic value. As discussed at the end of Section \ref{sec:NScomp},
this number does not coincide with the number of particles. In the mentioned examples there is only one independent degrees of freedom, and the reduction cannot occur. The prediction is therefore that in e.g. acoustic oscillations \cite{qtest} or superfluid flows \cite{Leggett1,Leggett2,He3,atomt} no deviation from the Schrodinger or standard wave equation can be observed, even for arbitrarily large number of particle involved.
Similar considerations apply for the experiments on the interference of "large" physical objects \cite{interference}.
The simplest possibility to witness the appearance of the reduction process would be to put a system
in a linear superposition of excited states at increasing number of independent degrees of freedom, e.g. by increasing the total excitation of the system. The model predicts that there should be an abrupt disruption of any quantum superposition as the number of independent degrees of freedom increases. Of course this is a quite vague and schematic suggestion. The realization of such a type of experiment, if possible, would require a separate study, which is well outside the scope of the paper.\par 
What has been only marginally touched along the presentation of the model is the question of the intrinsic nature of the wave function, e.g. if it is ontological or epistemic. As already discussed, the nonstandard component appears to have some ontological features, as a consequence of the formalism. This is anyhow a semi-philosophical question, and one can only say that the wave function appears as the simplest tool to describe an enormous variety of phenomena. Why we need such an extended object (at least in realistic model) remains of course an unanswered question. To this respect let us notice that the model does not introduce explicitly the particle degrees of freedom, like in Bohm' s Quantum Mechanics \cite{Kass}. All the physical systems are described by wave functions only. 
\vskip 0.2 cm
\par\noindent
\textbf{Acknowledgements.} 
\par
The author gratefully acknowledges Prof. L. Amico for introducing him to the Josephson persistent-current Qubits.

\section*{Appendix A. A schematic introduction to nonstandard analysis\label{nons}}
\indent
Despite in Quantum Mechanics the precision of the position of a particle cannot be arbitrarily accurate, because of the uncertainty principle, one assumes that the physical space-time is a real number manifold of suitable dimensions. However this can be considered a non necessary limitation. Nonstandard analysis has shown rigorously that the set of real numbers can be enriched by the introduction of additional numbers, usually indicated indeed as nonstandard numbers. 
In this section we give a very limited introduction to nonstandard analysis, in order to provide some explanations of  
of the terminology and summarize some basic properties of nonstandard analysis. The presentation is restricted
mainly to the features of nonstandard analysis which are used in the paper. The language is not in a rigorous mathematical style, but rather it appeals to intuition, as it is appropriate for a physicist reader. And of course no proof is presented.\par
A possible way of introducing nonstandard analysis is the axiomatic one \cite{Keisler,Diener,Nelson,Loeb,Benci}. Let us remind that the real axis is a complete ordered field, which means schematically that for its elements the operations of addition and multiplication are well defined and moreover a relation of ordering is given, i.e. for any pair of real numbers it is possible to establish which one is larger and eventually if they are equal. The completeness property
refers to the fact that any finite set in \textbf{R} has a lowest upper bound. Following ref. \cite{Keisler,Benci} we first introduce two axioms for the nonstandard real axis \textbf{R}$^*$
\par\noindent
A) \textbf{R}$^*$ is an ordered field, proper extension of \textbf{R}. \par\noindent
B) \textbf{R}$^*$ has a positive infinitesimal $\epsilon$, which has the property to be non-zero but smaller than any positive real r of \textbf{R}. 
\par\noindent The extension is a proper one, in the sense that there are elements belonging to
\textbf{R}$^*$ but not to \textbf{R}. Notice that \textbf{R}$^*$ is not complete. It turns out that any proper extension of \textbf{R} cannot be complete. This means that not all finite sets has a least upper bound. This is the case of the set of infinitesimals. If $\epsilon$ is an infinitesimal, $H\, =\, 1/\epsilon$, $\epsilon > 0$, is larger than any real number in  
\textbf{R}. Such a number will be indicated as unlimited. The set of unlimited numbers has no greatest lower bound belonging to \textbf{R}$^*$ or \textbf{R}.
It follows that the set of unlimited numbers are strictly disjoint from \textbf{R}.
From the physical point of view this means that all processes which can be considered as occurring within this set
can be assumed to be due to additional degrees of freedom with respect to the ones occurring in \textbf{R}. All that can be trivially extended to the negative sector of the real axis. Summarizing, an element $x$ of \textbf{R}$^*$ is infinitesimal if $ \vert x \vert < r $ , for any real $r$ of \textbf{R}, finite if $ \vert x \vert < r $, for some real $r$ of \textbf{R}, unlimited if $ \vert x \vert > r $ for any real $r$ of \textbf{R}. Notice that a finite element by definition is not infinitesimal. The expectations for the addition and multiplication
of two of the elements x,y in \textbf{R}$^*$ can be summarized as follows
\par
x,y infinitesimal , x+y and xy infinitesimal \par
x infinitesimal and y finite, x+y finite, xy infinitesimal \par
x,y finite, x+y and xy finite \par
x finite, y unlimited, x+y and xy unlimited \par
x,y unlimited, x+y and xy unlimited \par
x infinitesimal, y unlimited, x+y unlimited, xy not unique
\par\noindent
The last line can be understood from the following examples
\par
$\epsilon \ 1/\epsilon \,=\, $ finite \par
$\epsilon \ 1/\sqrt{\epsilon} \,=\, $ infinitesimal \par
$\epsilon \ 1/\epsilon^2 \,=\, $ unlimited  \par 
\noindent Each one of these products is between an infinitesimal and an unlimited number, but the result depends on the relative 'size' of the unlimited number. \par 
One of the key nonstandard object is the monad of a finite element. Two elements x,y of \textbf{R}$^*$ are said to be infinitely close if x - y is an infinitesimal, and we write x $\sim$ y. Then, given an element x of \textbf{R}$^*$,
the monad of x, monad(x), is formed by all elements infinitely close to x. In particular each element of \textbf{R}
has a monad. It turns out that any finite element x of \textbf{R}$^*$ can be written as x $=$ r $+ \epsilon$, where r is an element of \textbf{R}, i.e. each monad contains a real number r of \textbf{R}. One of the key property of the monads is that any two of them, if they are not disjoint, then they coincide, i.e. they cannot have elements in common except when they coincide. From these properties it follows that to each element x of \textbf{R}$^*$ one can associate 
a unique element r of \textbf{R}, which is called the standard part of x, and one writes 
\begin{equation}
x \,=\, r \,+\, \epsilon \ \rightarrow \ r \,=\,  st(x)
\end{equation} 
\noindent
For an unlimited number the standard part does not exist. This means that \textbf{R} is embedded properly in \textbf{R}$^*$, provided we identify the standard part of any number of \textbf{R}$^*$ with the corresponding number of \textbf{R}. As in the case of the set of unlimited numbers, the additional nonstandard numbers that do not coincide with elements of \textbf{R} can be viewed physically as additional degrees of freedom. \par 
To each element $x$ of \textbf{R}$^*$ one can associate also what is called a galaxy, which is the set of numbers $y$ such that $y\, -\,  x$ is finite and this set is indicated as gal($x$).\par 
To complete the extension of \textbf{R} to \textbf{R}$^*$ one needs two other axioms \cite{Keisler}. The first one has to do with functions, and it asserts that each standard function $f$ in \textbf{R} has an extension, called canonical extension, to a nonstandard function $^*f$. A nonstandard function (in one variable) is just a correspondence that assigns to each element of a set of nonstandard numbers a unique element of another set of nonstandard numbers. Generalization to functions of several variables is trivial. Once nonstandard functions have been introduced, various operations in \textbf{R} can be extended to \textbf{R}$^*$, noticeably the operation of differentiation and integration,
for details see e.g. \cite{Keisler,Diener,Nelson,Benci}.
The last axiom \cite{Keisler} is the so called Transfer Principle,
which connects, loosely speaking, the validity of a statement in \textbf{R} to the validity of the same statement in \textbf{R}$^*$, when the different elements of the statement have been extended. Since the principle will not be explicitly used, we refer the reader to the above quoted literature. Finally we notice that the extension of to \textbf{R}$^*$ entails that the set of natural numbers \textbf{N} can be extended to the nonstandard set \textbf{N}$^*$,
which include unlimited integer numbers. \par  
Another method to introduce nonstandard analysis is the construction of an explicit model for \textbf{R}$^*$, which will satisfy the axioms 
and eliminates any doubt on the existence of infinitesimals and unlimited numbers as introduced axiomatically.
This is developed in the celebrated paper and text book of Robinson \cite{Robinson1,Robinson2}. \par 
The method is similar to the one used in elementary analysis for introducing the field of real numbers from the field of rationals. We sketch her the procedure for future purpose. As it is well known, in standard analysis the Cauchy sequences of rational numbers can be used to define the real numbers. Furthermore, as the field of real numbers \textbf{R} is complete, a Cauchy sequence of real numbers is convergent to a real number. It is possible to construct a new field by considering the set of general sequences of standard real numbers, which is called 'ultraproduct'. It is easy to see that this a field if the arithmetic operation are defined through their application term by term of two sequences.
To obtain an ordered and well defined field however it is not possible to introduce the operation of comparison, i.e. larger or smaller, term by term. Furthermore one needs an equivalence relation within the set of sequences. For all that it is necessary to select the terms of the sequences for which the comparison and equivalence can be applied.
Such a selection is coherently performed by introducing what is called an 'ultrafilter', which indeed defines the set of terms that has to be used for the comparison. Since the terms of a sequence are labeled by an integer number, the ultrafilter is a family of subsets of the set of integer numbers \textbf{N}. Once this is done the field is completely ordered and one can identify each equivalence class of sequences with a number. This new field is embedding the standard real numbers. Each real number r can be simply identified with the sequence where all terms are equal to r. We are not going to define what is an ultrafilter, which can be chosen in many but equivalent ways, but just assume that it exists. Of course the new field \textbf{R}$^*$ contains numbers which have no correspondence in \textbf{R}, i.e. all the sequences not equivalent to the above mentioned constant one. It is then possible to identify infinitesimal and unlimited numbers. To exemplify the case of infinitesimal, let us consider sequence in \textbf{R} convergent to zero. If it turns out that the terms of the sequence that are smaller than ant given real number have indexes belonging to the ultrafilter, 
then the sequence defines an infinitesimal. In fact 
 the number in \textbf{R}$^*$ defined by the sequence is smaller than any real number. \par 
Within the model it is possible to extend in a constructive way many standard mathematical objects to \textbf{R}$^*$ and show that the above mentioned axioms are indeed satisfied.
We consider the case of the extension of a standard function $f(x)$, which  appears several times in the text.
To extend the definition of $f$ at a given nonstandard point $x^*$, we consider the sequence $\{ x_n \}$ which define
$x^*$, and the corresponding sequence $\{ f(x_n) \} $. The latter defines in general a nonstandard number $f^*$. The correspondence between $x^*$ and $f^*$ defines a nonstandard function $^*f$
\begin{equation}
 \{ x_n \} \ \rightarrow \  \{ f(x_n) \} \,=\, f^* \,\equiv\, ^*f(x^*) 
\end{equation}   
\noindent This extension will be referred to as 'canonical extension'.     
Vice versa, given a nonstandard function, one can introduce its standard part
\begin{equation}
^*f(x^*) \ \rightarrow \ st(^*f(st(x^*))) \,\equiv\, f(x) 
\end{equation} 
\noindent It has to be noticed that the standard part of a canonically extended function $^*f$ not necessarily
coincide with the original standard function $f$, i.e. the two operations are not the 'inverse' of each other. However, for smooth enough (finite) functions this is the case. For an extended exposition of the ultrafilter model and several applications we can refer to \cite{Goldblatt,Albeverio}
\par\noindent 
\par\noindent
\section*{Appendix B. Nonstandard wave equation and its solution. \label{sec:AppendixB}} 
\par 
Let us consider the free wave equation for spinless and massless particles
\beq
 \frac{\partial^2 W_j(t)}{\partial t^2} \,-\, c^2 \frac{1}{d^2} (W_{j+1} - 2 W_{j} + W_{j-1}) \,= 0
\label{eq:ww1}
\eeq  
\par\noindent
where $W_j(t) = W(t,x_j)$, being $x_j = jh = j/N$ the nonstandard $j$-point, and $c$ the wave phase velocity. Here $j$ runs from $-M +1$ to $M - 1$, for a total of $2M - 1$ equations, with $ M = N^2 $. The values of $W_{-M}$ and $W_M$ must be fixed by the boundary conditions. 
\par\noindent
Eq. (\ref{eq:ww1}) defines a band matrix $A$
\beq
A_{j,j'} \,=\, \frac{1}{h^2} (\delta_{j,j'-1} - 2\delta_{j,j'} + \delta_{j,j'+1} )
\label{eq:bbm}
\eeq  
\par\noindent
with $j,j'$ running from $-M+1$ to $M-1$.
Then the wave equation can be written for the vector $W \,=\, (W_{-M}, ............. W_{M})$
\beq
 \frac{\partial^2 W(t)}{\partial t^2} \,-\, c^2 A W \,= 0 
\eeq
\par\noindent
supplemented by the boundary conditions.
For any value of $M$ (finite or hyper-finite) the matrix $A$ can be diagonalized. In fact if we take 
\beq
 \phi_k(x_j) \,=\, \frac{1}{\sqrt{2M+1}} \exp(\imath kj\pi/M)  
\label{eq:sol}
\eeq       
\noindent 
one gets
\beq
\begin{array}{ll}
\big(A\phi_k)(x_j) &\,=\, \frac{1}{d^2} \frac{1}{\sqrt{2M+1}} \big[ \exp(\imath k(j+1)\pi/M) + \exp(\imath k(j-1)\pi/M) \\ \nonumber
\ &\, \\ \nonumber
\  &\ \ \ \ \ \ \ \ \ \ \ \ \ \ \ \ \ \ \ \ \ \,-\, 2\exp(\imath kj\pi/M) \big]  \\ \nonumber 
\ &\, \\ \nonumber
\, &\,=\, \frac{1}{d^2} \frac{1}{\sqrt{2M+1}}\exp(\imath kj\pi/M) \big[ \exp(\imath k\pi/M) + \exp(-\imath k\pi/M) - 2 \big]  \\ \nonumber
\ \\ \nonumber
\, &\,=\, - \frac{4}{d^2} \sin^2(k\pi/2M) \phi_k(x_j)
\end{array}
\eeq
\noindent Here $ k = 0, \pm 1, \pm 2, \cdots\cdots \pm M $. The eigenfrequencies are therefore
\beq
\omega(k) \,=\, \pm \frac{2}{d} \sin(k\pi/2M)
\eeq
\noindent and the solutions of the wave equation read
\beq
\phi_k(t,x_j) \,=\, \phi_k(x_j) \exp(\imath \omega(k) t)
\label{eq:phit}
\eeq
\par The solutions of Eq. (\ref{eq:sol}) correspond to periodic boundary conditions.
For massive particles the wave equation includes a mass term and Eq. (\ref{eq:ww1}) becomes the Klein-Gordon equation.
\noindent For the standard part of the w.f. the expression of the frequencies merges in the usual ones. Furthermore in the low wave number limit the Klein-Gordon equation becomes the Schr\"odinger equation. 
For particles with spin the Dirac equation and its ultra-relativistic limit have to be used, and the formalism can be easily extended to this case. \par 
All that can be generalized to a three dimensional space. 
The nonstandard plane wave for the wave number $\kk = (k_x,k_y,k_z)$ takes the form 
\beq 
\psi_k(t,\mathbf{j}) \,=\, \frac{1}{(2M+1)^{3/2}} \exp\Big(\imath(\frac{\kk \mathbf{j} \pi}{M} - \omega(\kk) t) \Big)
\label{eq:pw3}
\eeq
\noindent where the vector $\mathbf{j} = (j_x,j_y,j_z)$ labels the space position $\mathbf{x}_j$ in the three dimensional hyperfinite real space.
\beq
\mathbf{x}_j \,=\, d \mathbf{j} \ \ \ \ \ ; \ \ \ \ \ j_x, j_y, j_z \,=\, 0, \pm 1, \pm 2 \cdots\cdots N^2 
\eeq
\noindent The frequency $ \omega(\kk) $ is now given by
\beq
 \omega(\kk) \,=\, \frac{2}{d}\sqrt{\sin^2(k_x\pi/2M) + \sin^2(k_y\pi/2M + \sin^2(k_x\pi/2M)}  
\label{eq:freq3} 
\eeq
\noindent which is clearly anisotropic. \par 
The direction of propagation of a plane wave corresponds to the choice in each axis
of a multiple (standard or nonstandard) of the infinitesimal momentum step within the hyperfinite momentum lattice. 
All the possible directions at the standard level can be evenly covered by these nonstandard directions, so that anisotropy cannot be revealed at the standard level. Similar considerations apply in coordinate space, where the points of the planes perpendicular to the wave vector correspond to equal phase.  
\par 
Let us now consider a more general case of a single particle bound in a scalar potential well $ v(x) $.
This w.f. includes the standard component which is the solution $ \Psi_S(t,x) $ of the wave equation
\beq
\big(-\frac{\partial^2}{\partial t^2} \,+\, \Delta \,-\, (m + v(x))^2 \big) \Psi_S(t,x) \,=\, 0
\label{eq:swe}
\eeq 
\noindent and the nonstandard component , in connection with the extension in \Rns of the wave equation. In particular the Laplacian $ \Delta $ is defined in \Rns as in the text and above, Eq. (\ref{eq:ww1}). There is no matrix element of $v(x)$ between standard and nonstandard component, since they belong to different spaces of wave vectors. The choice of the nonstandard component, corresponding to a standard w.f. as given in Eq. (\ref{eq:ansatz}), includes its possible nonstandard singularities. In fact, at the reference time $ t = 0 $, the nonstandard component can be written   
\beq
\Psi_{NS}(0,x_j) \,=\, G(0,x_j) ^*\Psi_S(0,x_j)
\label{eq:GFN}
\eeq
\noindent where  $ ^*\Psi_S $ is the canonical extension of $ \Psi_S $ and 
\beq
 G(0,x_j) \,=\, \sum_k \exp(-\imath k x_j)
\label{eq:Gt0}\eeq
\noindent As specified in the main text, the nonstandard spectrum of wave vectors is a equally spaced set in the non linear region. The summation can then be performed
\beq
\Psi^{NS}(0,x_j) \,=\, \Psi^S(0,x_j) \exp(\imath p_2 x_j) \sin((p_1+\Delta p_2)x_j)/\sin(\Delta p_2 x_j)
\label{eq:sumes}\eeq
\noindent where $ \Delta p_2 = \Delta p/2 $, $ p_1 = l_0 \Delta p_2 $ and $ p_2 = p_0 + p_1 $. One gets a set of singularities, regularly spaced by $ \Delta x = 2\pi/\Delta p_2 $ and spatial width $ 2\pi/(p_1+\Delta p_2) $. The introduction of the nonstandard component is therefore equivalent to consider a set of (nonstandard) singularities of the wave function, beyond the usual assumption of smoothness in standard QM. The latter is demanded to avoid infinite energies, which is excluded in the model by the different meaning of the nonstandard component. The peak value of each singularity in absolute value equals $ l_0 + 1 $. 
At a generic time $ t $ the nonstandard component of a non free propagating wave function is
\beq
\Psi^{NS}(t,x_j) \,=\, \Psi^S(t,x_j) \sum_k \cos(\omega_k t) \exp(\imath k x_j) 
\label{eq:gent}\eeq
\par\noindent 
which includes only standing waves nonstandard components. The singularities are then time dependent, bu it is not possible to get a closed analytical expression for (\ref{eq:gent}).\par 
 The overlap between two states of the same system can be defined in the usual way, provided the integral is the nonstandard one. One gets for the nonstandard components of two w.f. 
\beq
\begin{array}{ll}
\, &< \Psi_{NS} \vert \Psi_{NS}' > \,=\, d \sum_{j r r'} \overline{\Psi}_{NS}(t,x_j) \Psi_{NS}(t,x_j)' \\ \nonumber
\ &\ \\ \nonumber 
\  &\,=\, \exp\big((\imath(k_r' - k_r) x_j\big) \cos(\omega_r t) \cos(\omega_r' t) ^*\overline{\Psi}_S(t,x_j) ^*\Psi_S(t,x_j)'  
\end{array}
\eeq
\noindent As discussed in the text, the difference of two distinct wave numbers is assumed to be unlimited. It follows that only the terms with $ k_r = k_r' $ survive, since the product of the two standard components (canonical extended) can contain only standard wave numbers. The summation over $ x_j $ then produces the overlap of the two standard components
\beq
\begin{array}{ll}
< \Psi_{NS} \vert \Psi_{NS}' > &\,=\, \sum_r \cos(\omega_r t)^2 \big[\, d \sum_j  ^*\overline{\Psi}_S(t,x_j) ^*\Psi_S(t,x_j)'\, \big] \\ \nonumber
\ &\  \\ \nonumber 
\ &\,=\,  \sum_r \cos(\omega_r t)^2 \, < \Psi_S \vert \Psi_S' >
\end{array}
\eeq
\par\noindent
Then the nonstandard part of the overlap is determined by the  standard one, which is one reason of the choice of Eqs.(\ref{eq:GFN},\ref{eq:Gt0},\ref{eq:gent}).\par 
As for the nonstandard wave equation, we notice that if we substitute Eq. (\ref{eq:gent}) in the equation, one gets necessarily unlimited terms of different orders, and the solution to all orders is too demanding. However with the expression (\ref{eq:gent}) 
the terms with the highest order $ N^2 $ vanish. In this sense we will consider that this expression solves the nonstandard wave equation. 
\section*{Appendix C. Matrix elements.\label{sec:AppendixC}}
\par 
In this Appendix we consider the interaction matrix elements between the nonstandard components of the 
states $l, m$,
defined and discussed in Section \ref{sec:fpro}. In the nonstandard sector the interaction involves unlimited wave numbers and frequencies and it must be infinitesimal. We schematize the interaction $ V^{NS} $ as
\beq
V^{NS}(x_i,x_j) \,=\, v_0 \delta(x_i - x_j)
\label{eq:vNS}\eeq
\noindent where $ v_0 $ is an infinitesimal strength and $\delta$ is the nonstandard version of the Dirac delta function.
We consider the case of particle detection.
To be specific, one can imagine that each one of the states $l, m$ is the result of the standard evolution of the initial state corresponding to a particle hitting a position detector, and that the state of the system particle+detector, before the reduction, is a linear combinations of them. For an ideal detector therefore they differ only for their mean position in the detector. \par 
The general form of the nonstandard matrix element can be written
\beq
< \Psi_l \vert V^{NS} \vert \Psi_m > \,=\, v_0 d\sum_j \rho_{NS}^{(l)}(t,x_j)\, \rho_{NS}^{(m)}(t,x_j)
\label{eq:gme}
\eeq  
\noindent where the $\rho_{NS}$'s are the density matrices corresponding to the states $l,m$. Their expression can be obtained from the generalization of Eq. (\ref{eq:ansatz}) to the NS wave function of a many-body system of A particles
\beq
\begin{array}{ll}
\ &\Psi_{NS}^{(l)}(t,x_{j_1},x_{j_2},\cdots\cdots,x_{j_A}) \,=\,  \\ \nonumber
\ &\  \\ \nonumber
\ &\ \ \ \ \ \sum_{r_1,r_2 \cdots r_A} \prod_n \cos(\omega_{r_n}t) \exp(\imath p_{r_n} x_{j_n})\, \phi_S^{(l)}(t,x_{j_1},x_{j_2},\cdots\cdots,x_{j_A})   
\end{array}
\label{eq:many}
\eeq 
\noindent where the $x_j$ run over the hyperfinite real axis, with $ j $ labeling its  position on it and the 
$ p_r $ run over the spectrum of the nonstandard wave vectors given in Eqs. (\ref{eq:array},\ref{eq:range1}). 
The corresponding density matrix is obtained by multiplying $ \Psi_{NS}^{l} $ by its complex conjugate and summing over A - 1 coordinates.
For simplicity we assume the w.f. to be symmetric or anti-symmetric under exchange of coordinates, so one can select 
$ x_{j_1} $ as the free coordinate. Taking into account the orthogonality between plane waves, one gets
\beq
\begin{array}{ll}
\rho_{NS}^{(l)}(t,x_j) &\,=\, \sum_{r_1,r_1'} \cos(\omega_{r_1}t)\cos(\omega_{r_1'}t) 
\exp(\imath(p_{r_1} - p_{r_1'})(x_j - X_l)\times \\ \nonumber 
\ &\ \\ \nonumber
\ &\ \ \ \ \times\sum_{r_2,\cdots,r_A} \prod_{n=2}^A \cos^2(\omega_{r_n}t)\, ^*\rho_S^{(l)}(t,x_j)
\end{array}
\label{eq:rnsmany}
\eeq
\noindent where $ ^*\rho_S $ is the canonical extension of the standard part of the density matrix, and $ X_l $ is the
average position of the state $ \Psi^{(l)} $. An essential property of this density matrix is its scaling with the number of particles. As shown below, the density matrix scales as $ \lambda^{A-2} $. This means that it is increasing exponentially with the number of degrees of freedom. This is at variance with the standard part of the density matrix, which can be normalized to 1 or to the number of particles.
\par 
Substituting these expression for the density matrices in Eq. (\ref{eq:gme}) and taking into account that any standard wave vector is strictly negligible with respect to an unlimited wave vector, one finally gets
\beq
\begin{array}{ll}
\!\!\!\!&< \Psi_l \vert V^{NS} \vert \Psi_m > \,=\, \\ \nonumber
\ &\  \\ \nonumber
\ &\ \ \ \ \ v_0 \sum_{\{r,s\}} \delta_{Kr}(r_1-s_1+s_2-r_2)
\exp(\imath(p_{r_1}-p_{s_1})(X_l-X_m)) \\ \nonumber
\ &\    \\ \nonumber 
\ &\ \ \ \ \ \times I^{(lm)} \cos(\omega_{r_1}t) \cos(\omega_{s_1}t) \cos(\omega_{r_2}t) \cos(\omega_{s_2}t)
(\sum_{n=2}^A \cos^2(\omega_{r_n}t))^{2(A-2)}
\end{array}
\label{eq:final}
\eeq   
\noindent where $ \delta_{Kr} $ is the discrete Kronecher delta function, extended to the nonstandard sector, and 
\beq
I^{(lm)} \,=\, \int dx \overline{\rho}_S^{(l)}(t,x) \rho_S^{(m)}(t,x)
\label{eq:intro}
\eeq
\noindent is the overlap integral of the two standard density matrices. In the wave number summation of Eq. (\ref{eq:final}) one has to exclude terms with equal momentum pairs, i.e. where both $ r_1 = r_2 $ and $ s_1 = s_2 $,
since these terms include the zero wave number and frequency component of the interaction. Possible frequency degeneracy
can introduce an extra factor in the summation of Eq. (\ref{eq:final}), but the structure of the matrix elements is the same and the conclusions would be the same.
\par
The expression of Eq. (\ref{eq:final}) has the structure of Eq. (\ref{eq:inme}) in the text, where, apart from trivial numerical factors
\beq
\begin{array}{ll}
 {\cal I}^{lm} &\,=\, v_0 I^{(lm)} (\sum_{n=1}^\lambda \cos^2(\omega_{r_n}t))^{2(A-2)} \,=\, \\ \nonumber
\ &\  \\ \nonumber
\ & v_0 I^{(lm)} \big[\frac{1}{2}  \big( \lambda \,+\, \sum_{n=1}^\lambda \cos(2\omega_{r_n}t)\big)\big]^{2(A-2)} \,= \\ \nonumber
\ &\ \\ \nonumber
\ & v_0 I^{(lm)} \big[ (\frac{1}{2}\lambda)^{2(A-2)} \,+\, R \big]
\end{array}
\label{eq:calI}\eeq 
\noindent
The summation over cosine functions in the parenthesis is expected to include with the same frequency positive and negative contributions. Then the summation is proportional at most to $ \sqrt{\lambda} $, which is strictly negligible with respect to $ \lambda $.
It follows that the matrix element is just a nonstandard oscillating functions with the set of frequencies
\beq
\Omega_s \,=\, \pm \omega_{r_1} \pm \omega_{s_1} \pm \omega_{r_2} \pm \omega_{s_2}
\label{eq:omegas}
\eeq 
\noindent if only positive values of the frequencies are included. All that complete the estimate of the matrix element,
which is then proportional to $ \lambda^{2(A-2)} $, times the oscillating function, in agreement with 
Eqs. (\ref{eq:rme}). \par 
As discussed in Sec. \ref{sec:fpro} a condition for the presence of a stochastic process is that the differences between 
any two frequencies of Eq. (\ref{eq:omegas}) is larger than a given unlimited value $\Delta \Omega $. A generic one of these frequencies, compatible with the conservation of wave numbers, can be written
\beq
\Omega \,=\, \omega(p_1 + \Delta p) - \omega(p_1) - ( \omega(p_2 + \Delta p) - \omega(p_2) )
\label{eq:gfreq}\eeq
\noindent with $p_1, p_2 $ two arbitrary wave numbers of the spectrum defined in Sec. \ref{sec:fpro}. The choice of the sign
has been selected to minimizing the absolute value of $ \Omega $ and then of the difference we are looking for.
For the same reason we take $ \Delta p $ equal to the step as defined in that spectrum. The closest frequency $ \Omega' $ is obtained by replacing $ \Delta p $ by $ 2\Delta p $. Then, after a set of trigonometric manipulations, one finds
\begin{eqnarray}
\Delta \Omega &\,=\, | \Omega' - \Omega | \,=\, \frac{8}{d} \sin^2(\frac{\Delta p}{4N}) \sin(\frac{p_1 + p_2 + \Delta p}{2N}) \\ \nonumber
\ &\  \\ \nonumber
\ \  &\,\sim\, \frac{\Delta p^2}{2N} \sin(\frac{p_1 + p_2 + \Delta p}{2N}) \
\label{eq:diff}\end{eqnarray}
\noindent In order that $ \Delta \Omega $ be unlimited one has to take $ \Delta p \sim N^\eta $, with $ \frac{1}{2} < \eta < 
 1 $, as specified in the text. This constraint can be considered also dictated by phenomenology.
\par 
Notice that it is essential that the frequency spectrum is non-linear in the wave vector. \par 
If the wave number spectrum is taken different for each coordinate one has to symmetrize among the different wave vectors in the nonstandard factor, or, equivalently, among the different coordinates. In this way the symmetry of the nonstandard component is the same as the standard one, i.e. symmetric or anti-symmetric. The expression of the matrix elements contains the summation over the exchange terms but the conclusions are the same. \par 


\end{document}